\documentclass[12pt]{article}

\usepackage{amssymb}
\usepackage{latexsym}
\usepackage{epsfig}

\usepackage{tcolorbox}
\usepackage{color,soul}

\definecolor{azure(colorwheel)}{rgb}{0.0, 0.5, 1.0}
\definecolor{DarkViolet}{RGB}{148,0,211}
\definecolor{MyDarkBlue}{rgb}{0,0.1,0.7}
\definecolor{DarkBlue}{RGB}{0,0,153}
\definecolor{amber}{rgb}{1.0, 0.49, 0.0}
\definecolor{amaranth}{rgb}{0.9, 0.17, 0.31}
\definecolor{nicered}{rgb}{0.7,0.1,0.1}
\definecolor{brown}{rgb}{0.5,0.1,0.1}
\definecolor{nicegreen}{rgb}{0.0,0.3,0.0}
\definecolor{tealgreen}{rgb}{0.0, 0.51, 0.5}

\usepackage{tabularx} % extra features for tabular environment
\usepackage{amsmath}  % improve math presentation
\usepackage{graphicx} % takes care of graphic including machinery
\usepackage{float}
\usepackage{cite} % takes care of citations
\usepackage[final]{hyperref} % adds hyper links inside the generated pdf file
\hypersetup{
	colorlinks=true,       % false: boxed links; true: colored links
	linkcolor=red,        % color of internal links
	citecolor=blue,        % color of links to bibliography
	filecolor=magenta,     % color of file links
	urlcolor=cyan         
	}

\usepackage{ulem}	
	
\usepackage{fancyhdr}
\usepackage{datetime}

\graphicspath{{images/}}

\newcommand{\pd}{\partial}

\newcommand{\mc}{\mathcal}
\newcommand{\tx}{\mathrm}

\newcommand{\newc}{\newcommand}

\newc{\bako}[1]{\textcolor{DarkViolet}{#1}} % bakop color

\relax
\renewcommand{\theequation}{\arabic{section}.\arabic{equation}}
\def\be{\begin{equation}}
\def\ee{\end{equation}}
\def\bea{\begin{eqnarray}}
\def\eea{\end{eqnarray}}
\newcommand\fverb{\setbox\pippobox=\hbox\bgroup\verb}
\newcommand\fverbdo{\egroup\medskip\noindent%
                        \fbox{\unhbox\pippobox}\ }
\newcommand\fverbit{\egroup\item[\fbox{\unhbox\pippobox}]}

\newcommand{\bear}{\begin{eqnarray}}

\newcommand{\eear}{\end{eqnarray}}
\newbox\pippobox

\def\a{\alpha}

\def\g{\gamma}

\def\f{\phi}

\def\sq
\def\y{\psi}

\usepackage{orcidlink}
\def\idbako{\orcidlink{0000-0002-3012-6144}}
\def\idchar{\orcidlink{0000-0002-5364-4753}}
\def\idkand{\orcidlink{0000-0002-3018-5558}}

\begin{document}

\begin{titlepage}

\vspace*{1cm}
\begin{center}
{\bf \Large Traversable wormholes in beyond Horndeski theories}
\bigskip \bigskip \medskip

{\bf A. Bakopoulos\idbako}$^{\,(a)}$\,\footnote{Email: a.bakop@uoi.gr} 
{\bf C. Charmousis\idchar}$^{\,(b)}$\,\footnote{Email: christos.charmousis@th.u-psud.fr} 
{\bf P. Kanti\idkand}$^{\,(c)}$\,\footnote{Email: pkanti@uoi.gr}

 \bigskip
$^{(a)}${\it Division of Applied Analysis, Department of Mathematics,\\
University of Patras, Rio Patras GR-26504, Greece}

$^{(b)}${\it Université Paris-Saclay, CNRS/IN2P3, IJCLab, 91405 Orsay, France.}

$^{(c)}${\it Division of Theoretical Physics, Department of Physics,\\
University of Ioannina, Ioannina GR-45110, Greece}

\bigskip \medskip

\today

\bigskip \medskip
{\bf {Abstract:}}
\end{center}

\noindent We construct a large class of explicit, asymptotically flat and regular wormhole solutions in higher order scalar tensor theories. The solutions are vacuum solutions of scalar tensor theory and no matter (exotic or regular) is introduced in order to support them. They are constructed via a general disformal transformation of a seed black hole solution. The seed solutions belong to a particular Horndeski theory which requires the presence of all extended Galileons and has a higher dimensional Lovelock origin. As a result, the resulting wormholes are always solutions of general beyond Horndeski theory. The particular class of wormholes we study are parametrised by their ADM mass and two coupling constants of the theory, one related to their higher dimensional Lovelock origin and one to the disformal transformation itself. The latter of the coupling constants affects the throat size of the wormhole solutions, thus giving them a compact or non-compact nature, as well as their properties. 

\bigskip \medskip

\end{titlepage}

\setcounter{page}{1}
%%%%%%%%%%%%%%%%%%%%%%%%%%%%%%%%%%%%%%%%%%%%%%%%%%%%%%%%%%%%%%%%%%%%%%

\def\g{\gamma}
\def\go{\g_{00}}
\def\gi{\g_{ii}}

 %%%%%%%%%% THIS IS IGNORED %%%%%%%%%%%
 
 \section{Introduction}

 In recent years there has been impressive quality and plethora of observational data concerning compact objects. Observations are up to now compatible with General Relativity (GR) neutron stars and black holes. The first breakthrough came with the discovery of gravitational wave emission of compact object binaries  (see in particular \cite{LIGOScientific:2017vwq}, \cite{LIGOScientific:2020zkf}, \cite{LIGOScientific:2021qlt}) of the LIGO and Virgo observatories. Gravity waves constitute a direct detector of gravitational physics. In addition, networks of radio telescopes, constituting a virtual earth size telescope, the Event Horizon Telescope (eg., \cite{EventHorizonTelescope:2020qrl}) have given a direct image of the supermassive black hole M87 in the center of our closest galaxy. Furthermore, the GRAVITY collaboration has measured to good accuracy the redhift and pericenter precession of the star S2 orbiting the central black hole of our host galaxy and will continue to do so, for S2 and neighboring orbiting stars, with improving accuracy. Additionally there are, for example X-ray, observations of pulsars with the NICER mission \cite{Raaijmakers:2019dks} aiming to extract the equation of state of neutron stars. 
 
 These are but a few of the ongoing experiments, and observational data will continue to accumulate in the forthcoming years. It is fair to say that we are living a revolutionary epoch in gravitational physics. It is quite natural then, as a result, that several questions or unresolved issues arise associated to this remarkable inflow of data. These can be questioned from within the framework of GR, but may be also extended beyond the realm of GR. For example, what is the nature and characteristics of the secondary observed compact object in \cite{LIGOScientific:2020zkf}? Is it a very small black hole or a very heavy neutron star? At the end of the day, is there a mass gap in-between neutron stars and black holes, and what is its value? If \cite{LIGOScientific:2020zkf} has observed a neutron star secondary, how can we explain its resulting stiff equation of state as compared to the observation of \cite{LIGOScientific:2017vwq}? What is the high density equation of state of neutron stars, and how fast can they  rotate before being destabilised? 
 
 One can go further, and question if the compact object is a GR solution or not, and how to go about measuring departure from GR. How are black holes modified in alternative theories of gravity? Do there exist compact objects other than the standard neutron stars and black holes, such as regular black holes (see for example \cite{Baake:2021jzv,Babichev:2020qpr,Kleihaus:2019rbg,Kleihaus:2020qwo} and refs within), 
 or horizon-less backgrounds like wormholes \cite{Einstein-Rosen} \cite{Misner-Wheeler} \cite{Wheeler:1957} \cite{Ellis} \cite{Bronnikov_0} \cite{MT} \cite{MTY} \cite{Visser}? How do these exotic objects compare to their GR black hole counterparts and how can we tell them apart (see for example  \cite{Damour:2007ap}, \cite{Cardoso:2016rao}, \cite{Dai:2019mse}, \cite{Simonetti:2020ivl}, \cite{Bambi:2021qfo}, \cite{Bronnikov:2021liv})? For example, the hypothesis that Active Galactic Nuclei (AGN's) are not supermassive black holes but rather throats of macroscopic  wormholes has been considered (see \cite{Piotrovich:2020vev}, \cite{Bambi:2013nla} and references within). Furthermore, theoretical work (see e.g. \cite{Damour:2007ap}) shows that if the gravitational redshift is very strong at the throat, the wormhole ``looks" very much like a black hole since for a faraway observer the coordinate time becomes extremely slow near the throat. In other words, there are cases where the throat behaves, in astrophysical terms, very much like an event horizon, a one-way membrane. 
 
 Wormholes \cite{Einstein-Rosen, Misner-Wheeler, Wheeler:1957, Ellis, Bronnikov_0, MT, MTY, Visser} are a classic example of an exotic compact object, since it is the playground of provocative effects in GR such as time travel, exotic matter etc. They pose indeed several theoretical problems in GR and beyond, and we may therefore ask, can these problems be lifted in modified or alternative theories of gravity? A large number of works have addressed the above issues and studied wormholes in a variety of contexts. The violation of energy conditions has thus been attributed to quantum effects \cite{Epstein:1965zza} \cite{Visser:1990wi} \cite{Maldacena:2018lmt} \cite{Kundu:2021nwp},  phantom fields \cite{Lobo:2005us} \cite{Bolokhov:2012kn}, $k-$essence fields \cite{Armendariz-Picon:2002gjc} or a Chaplygin fluid \cite{Lobo:2005vc}. At times, this violation (see also \cite{Visser:2003yf}) has been either minimised or even eliminated by the use of different class of metrics \cite{Gonzalez-Diaz:1996iea} \cite{Bronnikov:2009na} \cite{Bronnikov:2013zxa} \cite{Bronnikov:2018uje} or alternative geometric theories \cite{Boehmer:2012uyw} \cite{Hohmann:2013dra} \cite{Shaikh:2015oha} \cite{Bronnikov:2015pha} \cite{Bronnikov:2016xvj} \cite{Bronnikov:2019ugl}. The most intensely studied approach however is the one according to which the energy conditions are violated not by the actual matter but by an effective energy-momentum tensor arising in the context of a modified theory of gravity. These theories include $f(R)$ theories \cite{Lobo:2009ip} \cite{Bronnikov:2010tt} \cite{Harko:2013yb} \cite{Kuhfittig:2018vdg} \cite{Karakasis:2021tqx} \cite{Ghosh:2021dgm}, 
 extra dimensions \cite{Bronnikov:1996de} \cite{Bronnikov:1997gj} \cite{Mehdizadeh:2015jra},
 brane-world models \cite{Bronnikov:1997ba} \cite{Bronnikov:2002rn} \cite{Lobo:2007qi} \cite{Kar:2015lma} \cite{Bronnikov:2019sbx},
 non-minimally coupled scalar-tensor theories
  \cite{Garcia:2010xb} \cite{Sushkov:2011jh} \cite{Korolev:2014hwa} \cite{Korolev:2020ohi} \cite{Korolev:2020yyy}, or
  higher-derivative gravitational theories
 \cite{Kanti:2011jz, Kanti:2011yv, Shaikh:2017zfl, Antoniou:2019awm, Canate:2019spb, Ibadov:2020ajr, Brihaye:2020dgo, Ibadov:2020btp, Ibadov:2021oqf} to mention a few indicative works. The stability behaviour of the wormhole solutions emerging in the context of different theories has also undergone an intensive study over the years \cite{Bronnikov:2001ils} \cite{Lobo:2003xd} \cite{Bronnikov:2004ax} \cite{Bronnikov:2005an}, \cite{Bronnikov:2012ch} \cite{Bronnikov:2013coa} resulting at times to no-go theorems \cite{Bronnikov:2006pt} \cite{Evseev:2017jek} and forcing us to move to theories beyond GR for viable solutions.

 On the theoretical front, a good starting point for physically interesting solutions are indeed scalar tensor theories and secondly no hair theorems (or, more precisely how to evade them). The former provide a non trivial,  robust and measurable departure from GR with higher order derivative theories evading Ostrogradski ghosts :  namely Horndeski \cite{Horndeski:1974wa}, beyond Horndeski \cite{Gleyzes:2014dya} or DHOST theories \cite{Langlois:2015cwa}, \cite{Crisostomi:2016czh}. On the second front, classical no hair theorems concerning black holes \cite{Hui:2012qt}, \cite{Volkov:2016ehx}, \cite{Herdeiro:2015waa}, \cite{Babichev:2016rlq} (but also neutron stars \cite{Lehebel:2017fag} and wormholes \cite{Evseev:2017jek})  can be extended beyond GR and in particular to scalar tensor theories. Most of the attention has been given to black holes, where there are several ways to circumvent no hair theorems: as all theorems, they obey certain hypotheses which nevertheless can be broken (see for example \cite{Babichev:2013cya}, \cite{Sotiriou:2014pfa}, \cite{Antoniou:2017acq}). The subject of black holes in higher order scalar-tensor theories goes back quite a long way to EFT actions emanating from string theory such as \cite{Kanti:1995vq},\cite{Campbell:1990ai}, \cite{Chan:1995fr}, \cite{Charmousis:2009xr}. The nature of the solutions, explicit or numerical, turns out to be related to the presence of certain symmetries for the scalar field. 
 
 More recently, a new class of explicit black hole metrics have been constructed in theories where the scalar field has  shift and parity symmetry. A large subclass of these are dubbed stealth \cite{Babichev:2013cya}, \cite{Kobayashi:2014eva} in the sense that their metric is Ricci or Einstein flat, but, with a non trivial scalar field. Due to their stealth nature, they are less easy to distinguish from classical GR solutions. Recently however, using techniques such as disformal transformations, rotating black holes have been analytically constructed \cite{Anson:2020trg}, starting from a stealth Kerr solution \cite{Charmousis:2019vnf}. These disformed Kerr metrics are solutions to specific DHOST theories \cite{BenAchour:2020fgy} and are genuinely different from Kerr -- for example, they do not obey the no hair GR relation \cite{Anson:2021yli}. They have several distinguishable characteristics, which could be detectable in the near future. 
 
  On the other hand it is quite intriguing that once parity symmetry for the scalar field is broken, analytic solution methods seem no longer efficient. In fact, although solutions are known since a long time \cite{Kanti:1995vq}, \cite{Kanti:1997br}, they have been constructed numerically (see for example , \cite{Sotiriou:2014pfa},\cite{Antoniou:2017hxj}, \cite{Doneva:2017bvd}, \cite{Silva:2017uqg}, \cite{Babichev:2016fbg}, \cite{VanAelst:2019kku}, \cite{Bakopoulos:2019tvc}, \cite{Bakopoulos:2020dfg}) or perturbatively (see for example \cite{Campbell:1990ai}, \cite{Yunes:2011we}). Their non stealth nature on the other hand renders them very interesting from a phenomenological point of view. In particular a large and interesting class of these harbours the Gauss-Bonnet curvature scalar (see the discussion in \cite{Sotiriou:2014pfa}) which is a typical parity breaking term and is a natural term to expect from string theoretic or other fundamental (geometric) considerations. 
 
 Amongst this on-growing plethora of solutions there is one very recently found black hole \cite{Lu:2020iav}, that blends together properties coming from both classes of solutions whilst having quite unique characteristics. First of all, it is an explicit, non parity preserving, non-stealth solution (but close enough to Schwarzschild \cite{Clifton:2020xhc}, \cite{Charmousis:2021npl}). It involves all types of Horndeski scalar tensor terms, including the Gauss Bonnet term and is asymptotically flat. The underlying scalar tensor theory is a subclass of Horndeski \cite{Horndeski:1974wa} and the solution was originally found  by Lu and Pang \cite{Lu:2020iav}, carefully taking a singular limit \cite{Glavan:2019inb}, of a Kaluza Klein reduction of a higher order Lovelock theory \cite{Charmousis:2012dw}, \cite{Charmousis:2014mia}. Shortly after, Hennigar et al \cite{Hennigar:2020lsl}, constructed the same theory by generalising a 2 dimensional construction originating from GR \cite{Mann:1992ar}, while Fernandes interestingly found that the theory in question has a conformally coupled scalar field \cite{Fernandes:2021dsb}. Neutron stars were then constructed presenting several intriguing properties \cite{Charmousis:2021npl} such as the absence of the previously mentioned mass gap in between black holes and neutron stars within this framework.
 
 In this paper starting from the a-fore mentioned black hole, we will construct a rather large class of analytic wormhole solutions. Previous efforts \cite{Kanti:2011yv}, \cite{Kanti:2011jz}, \cite{Antoniou:2019awm} have given numerical solutions of wormholes within Horndeski theory (where there are instability issues \cite{Evseev:2017jek}, \cite{Cuyubamba:2018jdl}, \cite{Rubakov:2015gza}) and needed, due to regularity, the presence of some distributional source of ordinary matter. In contrast, we will show that the explicit solutions found here will be everywhere regular supported solely from the scalar tensor gravity theory with no additional matter (exotic or not). Resulting from a disformal transformation of a solution arising in Horndeski theory they will be exact solutions of beyond Horndeski theory. 
 They will be parametrised (apart from their ADM mass) by two coupling constants relating them to the seed Horndeski and the beyond Horndeski theory. One is the coupling constant associated to their seed black hole or in a looser sense to their higher order Lovelock origin. The latter will be associated to the beyond Horndeski theory and will crucially parametrise their throat size and essential wormhole properties. As such, wormholes with a throat size comparable to their seed black hole horizon radius will constitute very compact objects similar to black holes. Alternatively, solutions with a large throat will be far less `compact' and will be characterised by small tidal forces and acceleration effects for light or test particles.

 In the next section we will present the Horndeski and beyond Horndeski theory and its spherically symmetric equations. We will furthermore consider disformal transformations and find how the Horndeski functionals change to the beyond Horndeski ones for the case of spherically symmetric and static spacetimes. Starting from the Lu-Pang black hole \cite{Lu:2020iav} we will then construct analytic wormhole solutions in section 3. We will establish that these are regular and traversable by particles and light in section 4. We will then discuss some of their important physical properties in section 5 and conclude in section 6.  
 
 %%%%%%%%%%%%%%%%%%%%%%%%%%%%%%%%%
 
\section{From Horndeski to beyond with disformal transformations}

We shall consider shift-symmetric Horndeski theory and beyond Horndeski theory. The former is parametrised by four functions $\{G_i: i=2,..,5\}=\{G_2,G_3,G_4,G_5\}$ which are functions of the scalar field $\phi$ only through its kinetic density $X=-\frac{1}{2}\,\pd_\mu\phi\,\pd^\mu\phi$. Its action functional is given by the expression
\be
\label{eq:Hfr}
S_\tx{H} = \displaystyle\int \tx{d}^4x \sqrt{-g} \,\left(\mc{L}_2+\mc{L}_3+\mc{L}_4+\mc{L}_5\right),
\ee
with
\begin{align}
\mc{L}_2 &=G_2(X) ,
\label{eq:L2fr}
\\
\mc{L}_3 &=-G_3(X) \,\Box \phi ,
\label{eq:L3fr}
\\
\mc{L}_4 &= G_4(X) R + G_{4X} \left[ (\Box \phi)^2 -\nabla_\mu\pd_\nu\phi \,\nabla^\mu\pd^\nu\phi\right] ,
\label{eq:L4fr}
\\
\begin{split}
\mc{L}_5 &= G_5(X) G_{\mu\nu}\nabla^\mu \pd^\nu \phi - \frac{1}{6}\, G_{5X} \big[ (\Box \phi)^3 - 3\,\Box \phi\, \nabla_\mu\pd_\nu\phi\,\nabla^\mu\pd^\nu\phi
\\
&\quad + 2\,\nabla_\mu\pd_\nu\phi\, \nabla^\nu\pd^\rho\phi\, \nabla_\rho\pd^\mu\phi \big].
\label{eq:L5fr}
\end{split}
\end{align}
The latter, the shift-symmetric beyond Horndeski theory, is given by two additional higher-order terms,
\begin{align}
\mathcal{L}^{\rm bH}_4&=F_4(X)\,\varepsilon^{\mu\nu\rho\sigma}\,\varepsilon^{\alpha\beta\gamma}_{\,\,\,\,\,\,\,\,\,\,\,\sigma}\,\partial_\mu\phi\,\partial_\alpha\phi\,\nabla_\nu\partial_\beta\phi\,\nabla_\rho\partial_\gamma\phi,\\[3mm]
\mathcal{L}^{\rm bH}_5&=F_5(X)\,\varepsilon^{\mu\nu\rho\sigma}\,\varepsilon^{\alpha\beta\gamma\delta}\,\partial_\mu\phi\,\partial_\alpha\phi\,\nabla_\nu\partial_\beta\phi\,\nabla_\rho\partial_\gamma\phi\,\nabla_\sigma\partial_\delta\phi,
\end{align}
parametrised by the functions $F_4$ and $F_5$. These are not independent of the Horndeski functions $\{G_4, G_5\}$; in fact,  the following relation holds in order to avoid an Ostrogradski ghost degree of freedom \cite{BenAchour:2016fzp},
\begin{equation}
    X G_{5X} F_4= 3 F_5 (G_4-2X G_{4X})\,,\label{BH}
\end{equation}
where a subscript in $X$ denotes a derivative with respect to $X$. The above theories depend solely on the derivatives of the scalar field and as such are manifestly shift symmetric.

We will focus on static and spherically-symmetric metrics,
\be
ds^2=-h(r)\,dt^2 + \frac{dr^2}{f(r)} +r^2 d\Omega^2\,
\ee
with a static scalar field, $\phi=\phi(r)$.  The independent field equations under these symmetries can be obtained in all generality (see for example the Appendix of \cite{Lehebel:2018zga} modulo simple typos).  For a start, the $(tt)$ gravitational equation does not involve $h$ and reads
\be
\begin{split}
\label{eq:Ettb1}
&G_2 +f\phi'X'G_{3X}+ \dfrac{2}{r} \left(\dfrac{1 - f}{r} - f'\right) G_4 + \dfrac{4}{r}f \left(\dfrac{1}{r} + \dfrac{X'}{X} + \dfrac{f'}{f}\right) X G_{4X}
\\
& + \dfrac{8}{r} fX X' G_{4XX}+ \dfrac{1}{r^2}f\phi'\left[(1-3f)\dfrac{X'}{X}-2f'\right]XG_{5X}-\dfrac{2}{r^2}f^2\phi'XX'G_{5XX}
\\
& +\frac{16}{r}f X^2 X' F_{4X}+ \frac{8}{r} f \left(  4 \frac{X'}{X}+\frac{f'}{f}+\frac{1}{r} \right)X^2 F_4
\\
&+\dfrac{12}{r^2}f^2\phi'X^2\left(\dfrac{2f'}{f}+\dfrac{5X'}{X}\right)F_5+\dfrac{24}{r^2}f^2\phi'X^2X'F_{5X}=0.
\end{split}
\ee
The radial component of the current $J^r$ associated to shift symmetry is given by:
\be
\begin{split}
\label{eq:staticJrbHb1}
J^r&=-f \phi' G_{2X} - f\dfrac{r h' + 4 h}{r h} X G_{3X} + 2 f \phi' \dfrac{f h - h + r f h'}{r^2 h}G_{4X} 
\\
&\quad+ 4 f^2 \phi' \dfrac{h +rh' }{r^2h} X G_{4XX} -f h' \dfrac{1 - 3 f}{r^2h} X G_{5X} +2 \dfrac{h' f^2}{r^2h} X^2 G_{5XX}\\
&\quad+8f^2 \phi\frac{h+rh'}{r^2h}X(2F_{4}+XF_{4X})- 12 \dfrac{f^2 h'}{r^2 h} X^2 (5 F_5+2 X F_{5X})
.\end{split}
\ee
This expression should in general be equated to zero if the primary scalar charge is set to zero; we will consider such a solution here (for a full discussion on no hair arguments involving spherical symmetry see \cite{Babichev:2017guv}). Either way $J^r=0$, solves the scalar-field equation for spherical symmetry, which can be written as  $\nabla_\mu J^\mu=0$ due to shift symmetry.  

Finally, rather  than the $(rr)$ gravitational equation itself, we use a linear combination with the $J^r=0$ equation as undertaken in \cite{Babichev:2016kdt}. Namely, if $\mc{E}^{rr}$ stands for the $(rr)$ gravitational equation, we write $\mc{E}^{rr}-J^r\pd^r\phi$. This combination should also be equal to zero and leads to a simpler equation:
\be
\begin{split}
\label{eq:Errb1}
&G_2 - \dfrac{2}{r^2h}(frh'+fh-h) G_4 + \dfrac{4f}{r^2h} (rh'+h) X G_{4X} - \dfrac{2}{r^2h}f^2h'\phi'XG_{5X}
\\
&+ \frac{8f}{r^2h}(rh'+h)X^2F_4+\dfrac{24}{r^2h}f^2h'\phi'X^2F_5 = 0.
\end{split}
\ee
Interestingly, the cubic Horndeski term $G_3$ disappears from this combination \cite{Lehebel:2018zga}. In summary, the three latter quite lengthy expressions (equated to zero) give the possible spherically symmetric solutions of beyond Horndeski theories in the absence of primary scalar hair.

The third crucial ingredient we will need are disformal transformations. 
It is well known that a disformal transformation of the spacetime metric tensor, involving some function $D=D(X)$, takes a seed Horndeski solution to a beyond Horndeski solution (see for example \cite{Zumalacarregui:2013pma}, \cite{Crisostomi:2016tcp}, \cite{BenAchour:2016cay}, \cite{BenAchour:2016fzp}). Let us consider such a case, where by barred quantities we will be henceforth denoting the seed ``known" Horndeski solution. As such we have $\bar{\phi}, \bar{h}, \bar{f}$ and of course $\bar{X}=-\frac{1}{2}\bar{f}\bar{\phi'^2}$ for some specific  $\{\bar{G_i}, i=2,3,4,5\}$ Horndeski theory. Then, via the disformal transformation that reads
\be
\label{disformal}
g_{\mu \nu}=\bar{g}_{\mu \nu}-D(\bar{X})\; \nabla_\mu \bar \phi \,\nabla_\nu \bar \phi\,,
\ee
we go to a new  ``image'' metric tensor which is a solution of the beyond Horndeski theory \{$G_i, F_4, F_5$\} with the same scalar field as solution. Given that $\phi$ is only a function of $r$, we have immediately that $\phi =\bar{\phi}$, $h=\bar{h}$ whereas the only quantities that do change for the image solution are,
\begin{equation}
f=\frac{\bar{f}}{1+2 D\; \bar{X}},\qquad X=\frac{\bar{X}}{1+2 D\; \bar{X}}\,. \label{fxd}
\end{equation}
We emphasize that by $D$ we mean a generic function of $X$ (or equivalently $\bar{X}$).

As we vary the ``disformability" function $D$, we span all values of the metric function $f$ for the given same metric function $h$ and scalar field $\phi$, while at the same time, we change theory according to specific rules established in \cite{Crisostomi:2016tcp}. For the needs of our analysis, though, it is important to express these rules in terms of the initial known barred (seed) solution. Using \cite{Crisostomi:2016tcp}, making use of the chain rule and adopting to our conventions\,\footnote{In our analysis, we have defined $X\equiv -f \phi'^2/2$ while in  \cite{Crisostomi:2016tcp} it holds that $X\equiv f \phi'^2$.}, the transformation rules read
\begin{eqnarray}
G_4&=&\frac{\bar{G_4}}{(1+2 \bar{X} D)^{1/2}}\,,\label{g4d}\\[1mm]
G_{5X}&=&\frac{\bar{G}_{5\bar{X}}(1+2 \bar{X} D)^{5/2}}{1-2 \bar{X}^2 D_{\bar{X}}}\,,\label{g5d}\\
F_4&=&(\bar{G_4}-2\bar{X}\bar{G}_{4\bar{X}})\frac{D_{\bar{X}}(1+2 \bar{X} D)^{5/2}}{2(1-2 \bar{X}^2 D_{\bar{X}})}\,,\label{f4d}\\
F_5&=&\bar{X}\bar{G}_{5\bar{X}}\frac{D_{\bar{X}}(1+2 \bar{X} D)^{7/2}}{6(1-2 \bar{X}^2 D_{\bar{X}})}\,.\label{f5d}
\end{eqnarray}
In the above, we have also used the relation
\begin{equation}
    D_X=D_{\bar{X}} \frac{(1+2 \bar{X} D)^2}{1-2 \bar{X}^2 D_{\bar{X}}}\,,
\end{equation}
while the constraint (\ref{BH}) is verified.
Our seed solution will make use of all functions of Horndeski theory, we therefore still need to figure out how $G_2$ and $G_3$ transform. To do this we use the fact that the disformed metric is also a solution of the field equations. Indeed, applying the above rules for $G_4,G_5,F_4,F_5$ in the field equations for spherical symmetry, we find in addition that $G_2$ and $G_3$ change according to the following rules \cite{Bakopoulos:2022csr}
\begin{eqnarray}
G_2=\frac{\bar{G_2}}{(1+2 \bar{X} D)^{1/2}}\,,\label{g2d}\\
G_{3X}=\bar{G}_{3\bar{X}}\frac{(1+2 \bar{X} D)^{5/2}}{1-2 \bar{X}^2 D_{\bar{X}}}\,.\label{g3d}
\end{eqnarray}

In summary, if we start from a given seed solution of Horndeski theory $\{\bar{G}_i\}$, then a disformal transformation of the form Eq. (\ref{disformal}) for a given function $D$ can take us to a solution of beyond Horndeski theory $\{G_i,F_4, F_5\}$. In the next section, starting from a black hole of Horndeski theory we will use this simple mapping to obtain a wormhole solution of beyond Horndeski theory (see also very recently \cite{Faraoni:2021gdl}, \cite{Chatzifotis:2021hpg}).

%%%%%%%%%%%%%%%%%%%%%%%%%%%%%%%%%%%%%%%%%%%%%%%%%%%%%%%%%%%%%%%%%%%%%%%%%%%%%%%%%%%%%%%

\section{Constructing a wormhole from a black hole}
\setcounter{equation}{0}

In order to construct a wormhole in beyond Horndeski theory, we will employ the disformal transformations on a specific Horndeski solution. As seed we use the solution found by Lu-Pang  \cite{Lu:2020iav} (see also \cite{Hennigar:2020lsl}, \cite{Fernandes:2020nbq}) described by the following theory functions of Horndeski, 
$$G_2=8\alpha \bar{X}^2,\; G_3=-8\alpha \bar{X},\; G_4=1+4\alpha \bar{X},\; G_5=-4\alpha \ln\bar{|X|}.$$
The solution reads \cite{Lu:2020iav}, 
%%%
\begin{equation}
   \bar{h}(r)=\bar{f}(r)=1+\frac{r^2}{2 \a}\left(1\pm\sqrt{1+\frac{8\a M}{r^3}}\right), \quad {\rm and} \quad \bar\f'=\frac{\sqrt{\bar{h}}\pm 1}{r \sqrt{\bar{h}}},\label{phih}
\end{equation}
%%%
where the prime denotes differentiation with respect to the radial coordinate $r$.
In what follows, we restrict our analysis to the asymptotically flat $(\bar{h}_{-},\,\bar{\f}_{-})$ branch. For this choice, the above solution describes a black hole with $M$ being the black-hole mass and $\alpha$ the constant coupling parameter of the theory. The spacetime geometry is characterised by the roots of $\bar h$, located at $r_{\pm}=M\pm \sqrt{M^2-\a}$, with the largest one being the event horizon, $r_h=r_+$, whereby $\alpha\leq M^2$. 
Note that, strictly speaking, the black hole inner horizon is ill defined as the scalar becomes imaginary in the interior of the outer horizon; this caveat can be remedied by introducing linear time dependence \cite{Charmousis:2021npl} something that will not be important however for our construction as the horizon region will be excised from spacetime. It is important however for our seed solution that $\bar X$ is {\it{not}} a constant so as that the disformal function $D$ is crucially  {\it{not trivial}}. 

Indeed, under the disformal transformation \eqref{disformal}, the metric functions and scalar field become: 
\be
     h = \bar h\,, \qquad
     f=\frac{\bar h}{1+ 2D(\bar X) \bar X} \equiv h \,W(\bar X)^{-1}\,, \qquad
      \f = \bar \f\,, \label{fun-dis}
\ee
where we have defined for convenience the function $W(\bar X) \equiv  1+ 2D(\bar X) \bar X$. The above functions, are solution to a beyond Horndeski theory, given by (\ref{g4d}-\ref{g3d}) and parametrized by $D$.
The new line-element of spacetime therefore reads 
%%%
\begin{equation}
   ds^2= - h(r)\,dt^2 + \frac{dr^2}{h(r)\,W^{-1}}+ r^2\,(d\theta^2 + \sin^2 \theta\,d \varphi^2)\,.
   \label{wormhole}
\end{equation}
%%%%%
If $W^{-1}$ never vanishes, then, $g^{rr}$ becomes zero only when $h(r)$ does. As a result, the new (beyond-Horndeski) solution describes again a black hole with its horizon located at the horizon of the initial Horndeski solution, i.e. at $h(r_h)=0$. In order to construct instead a wormhole geometry, we demand that $W^{-1}$ has a root, i.e.
%%%%%
\begin{equation}
    W(\bar X)^{-1}=0|_{r=r_0},
\end{equation}
%%%%%%%
at a radial coordinate value $r_0$ {\it larger} than $r_h$. Then, the {\it red-shift} function $h(r)$ remains finite and non-zero in the entire region $[r_0, \infty)$, while the {\it shape} function $f(r)=h W^{-1}$ vanishes at a single point $r=r_0$, where the wormhole throat resides \cite{MT}. Note therefore, that in this case, the event horizon at $r=r_h$ is cut out of spacetime as we have that $r\geq r_0$ and furthermore the Killing vector $\partial_t$ is globally (and not locally) static.

There is clearly a variety of choices that one could make for the disformal transformation function $W^{-1}$.  This function should however obey two additional constraints: (i) $W^{-1}>0$ for $r>r_0$ in order to preserve the spacetime signature, and (ii) $W^{-1}$ should reduce to unity at radial infinity for an asymptotically-flat spacetime.  A simple choice that respects all of the above conditions is the following,
%%%%%%
\be
W(\bar X)^{-1}=1-\frac{r_0}{\lambda}\,\sqrt{-2\bar X}\,,
\ee
%%%%%%%%%
where $\lambda$ is a positive dimensionless constant and the ratio $b_1=r_0/\lambda$ of dimension length, parametrizes our theory and in particular the disformal transform.  Taking into account the expression \eqref{phih} for $\bar{\phi}'$ in Horndeski theory, the $\bar X$ function may be written as
%%%%%
\begin{equation}
    \bar X = - \frac{1}{2}\,\bar h \bar{\f}'^2=-\frac{1}{2}\,\frac{(\sqrt{\bar{h}}-1)^2}{r^2}\,. 
\end{equation}
%%%%%%%%%% 
Then, the disformal transformation function becomes: 
%%%%%
\begin{equation}
    W(\bar X)^{-1}=
    %1-\frac{r_0}{\lambda}\sqrt{-2\bar X}=1 -\frac{r_0}{\lambda r}\left| \sqrt{h}-1 \right|=
    1-\frac{r_0}{\lambda  r}\left( 1-\sqrt{h}  \right). \label{G}
\end{equation}
%%%%%
In the above, we used the property that $0 < h \leq 1$ for $r \geq r_0$. Since, at the throat, we have $W(\bar X(r_0))^{-1}=0$, the above readily leads to the condition
%%%%%%
\begin{equation}
    h(r_0)=(1-\lambda)^2\,.
    \label{con_lambda}
\end{equation}
%%%%%%%%%%
In order to have a wormhole throat, the above equation constrains $\lambda$ to be in the range $0 < \lambda < 1$: for $\lambda = 0$, we find $h(r_0)=1$  which means that $r_0 \rightarrow \infty$, and the wormhole throat is pushed to infinity;  for $\lambda=1$, we obtain $h(r_0)=0$, therefore $r_0=r_h$, and the wormhole throat coincides with the black-hole horizon. In other words, as $\lambda\longrightarrow 1$, the observer time $t$ seems to be frozen as $g_{tt}\longrightarrow 0$. In this region of parameter space we expect the wormhole to be very similar or even indistinguishable from a black hole geometry (see the nice analysis in \cite{Damour:2007ap}). 

Solving Eq. \eqref{con_lambda}, we may explicitly obtain the radius of the throat in terms of the parameters $(\alpha, \lambda)$  and $M$, our mass integration constant{\footnote{One could readily redefine $\alpha$ and $M$ so as to get rid of $\lambda$ in \eqref{roots} but not without spoiling the ADM mass interpretation of $M$.}}, as
%%%%%
\begin{equation}
    r_0=\frac{M+ \sqrt{M^2- \alpha \lambda^3\,(2-\lambda)^3}}{\lambda(2-\lambda)}\,.
    \label{roots}
\end{equation}
%%%%%
Therefore $r_0$ fixes the size of the throat while the parameter $\lambda$ the gravitational redshift at the throat.
We will consider $\alpha \geq 0$ in our analysis following the generic constraints of the seed theory \cite{Clifton:2020xhc}, \cite{Charmousis:2021npl}. We see therefore that the throat size is maximal in $\alpha$ for  $\alpha=0$ which is the GR limit in the seed solution. We will come back to this interesting limit in a moment. Once $\lambda$ and $\alpha$ for the theory are fixed accordingly{\footnote{Although we will be using the $\lambda$ parameter for practical reasons, the theory parameters are in fact $\alpha$ and the ratio $b_1=r_0/\lambda$. This is because $r_0$ and $\lambda$ depend individually  on the mass parameter $M$ and therefore on the solution. One can then show that fixing the theory, $\alpha$ and $b_1$, for each $M$ satisfying \eqref{alpha_bound} there is a unique parameter $0\leq\lambda\leq 1$.}}, for different mass parameter $M$, we have a different size throat for the wormhole solution. 
According to Eq. \eqref{roots}, wormhole solutions exist for
%%%%%
\begin{equation}
    \alpha\leq\frac{M^2}{\lambda^3\,(2-\lambda)^3}. \label{alpha_bound}
\end{equation}
%%%%%%%%
%t later.
Note in passing that when $M=0$ the solution is just flat spacetime.
In summary, for a large enough ADM mass $M$ such that for given  $\lambda\in[0,1]$ and $\alpha$,  \eqref{alpha_bound} is satisfied, our seed black hole ($\alpha<M^2$) of Horndeski theory is transformed into a wormhole of throat (\ref{roots}) in beyond Horndeski theory.

% read,
A simple working example solution can be obtained, closest to GR, by considering the expansion at $\alpha\rightarrow 0$. In this case the seed metric is approximated by a GR Schwarzschild metric. The throat resides at $r_0\simeq \frac{2M}{\lambda (2-\lambda)}$ while the metric functions read,
\begin{equation}
   h(r)=1-\frac{2M}{r} +\mathcal{O}(\alpha), \quad f(r)=\left(1-\frac{2M}{r}\right)\left[1-\frac{2M}{\lambda^2(2-\lambda)r}\left(1-\sqrt{1-\frac{2M}{r}}\right)\right]+\mathcal{O}(\alpha).
   \label{wormhole0}
\end{equation}
%Interestingly. 
We will be making use of this simple enough example throughout our analysis in the forthcoming sections{\footnote{Note that although the $\alpha=0$ limit exists by continuity, we nevertheless insert the $\mathcal{O}(\alpha)$ notation as the equations of motion leave the scalar field undetermined at this point.} }

%%%%%%%%%%%%%%%%%%%%%%%%%%
\begin{figure}[t] 
%\lbfig{Fig_phi} 
\begin{center}
\hspace{0.0cm} \hspace{-0.5cm}
\includegraphics[height=.21\textheight, angle =0]{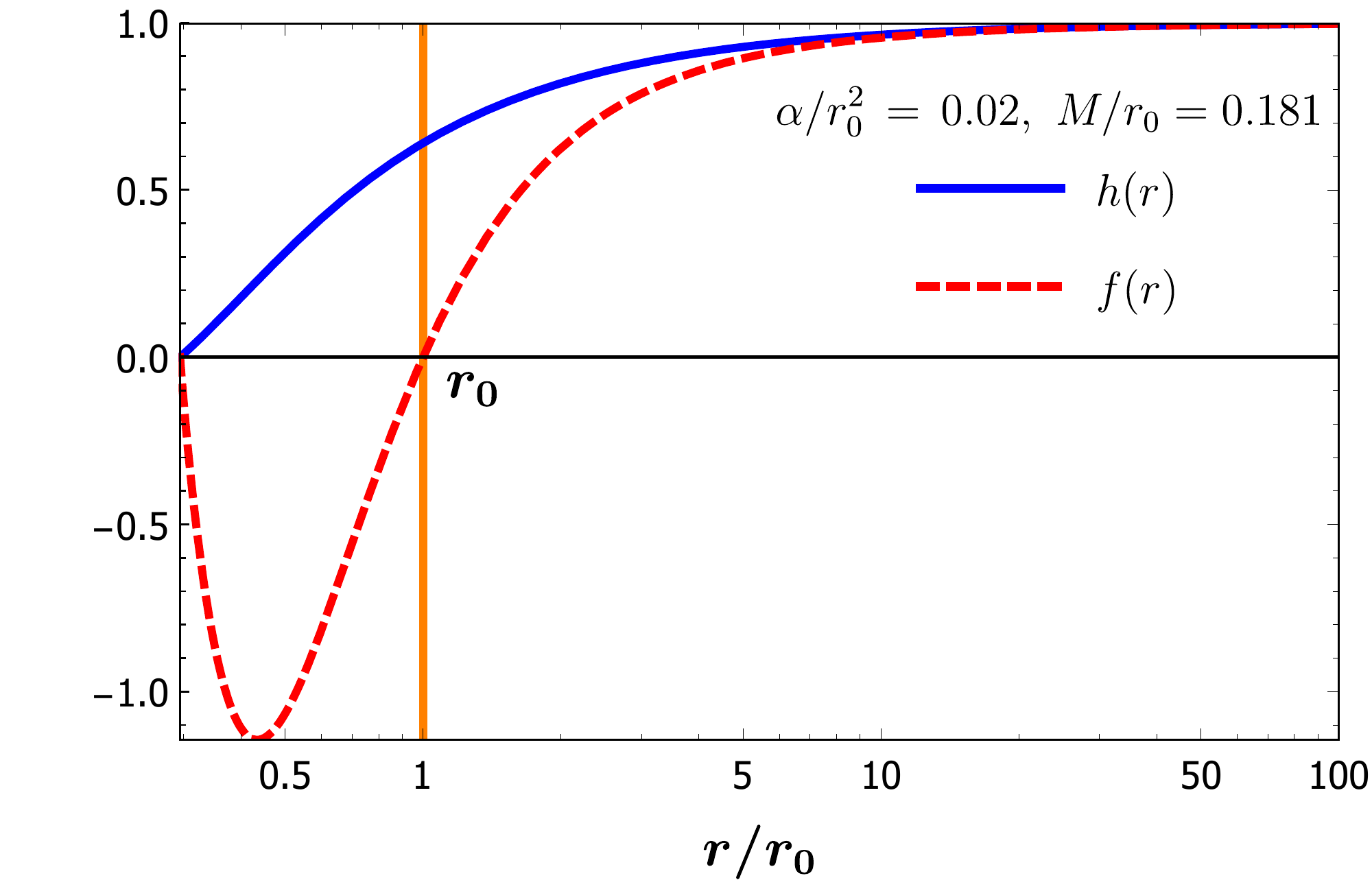}
\hspace{0.42cm} \hspace{-0.4cm}
\includegraphics[height=.21\textheight, angle =0]{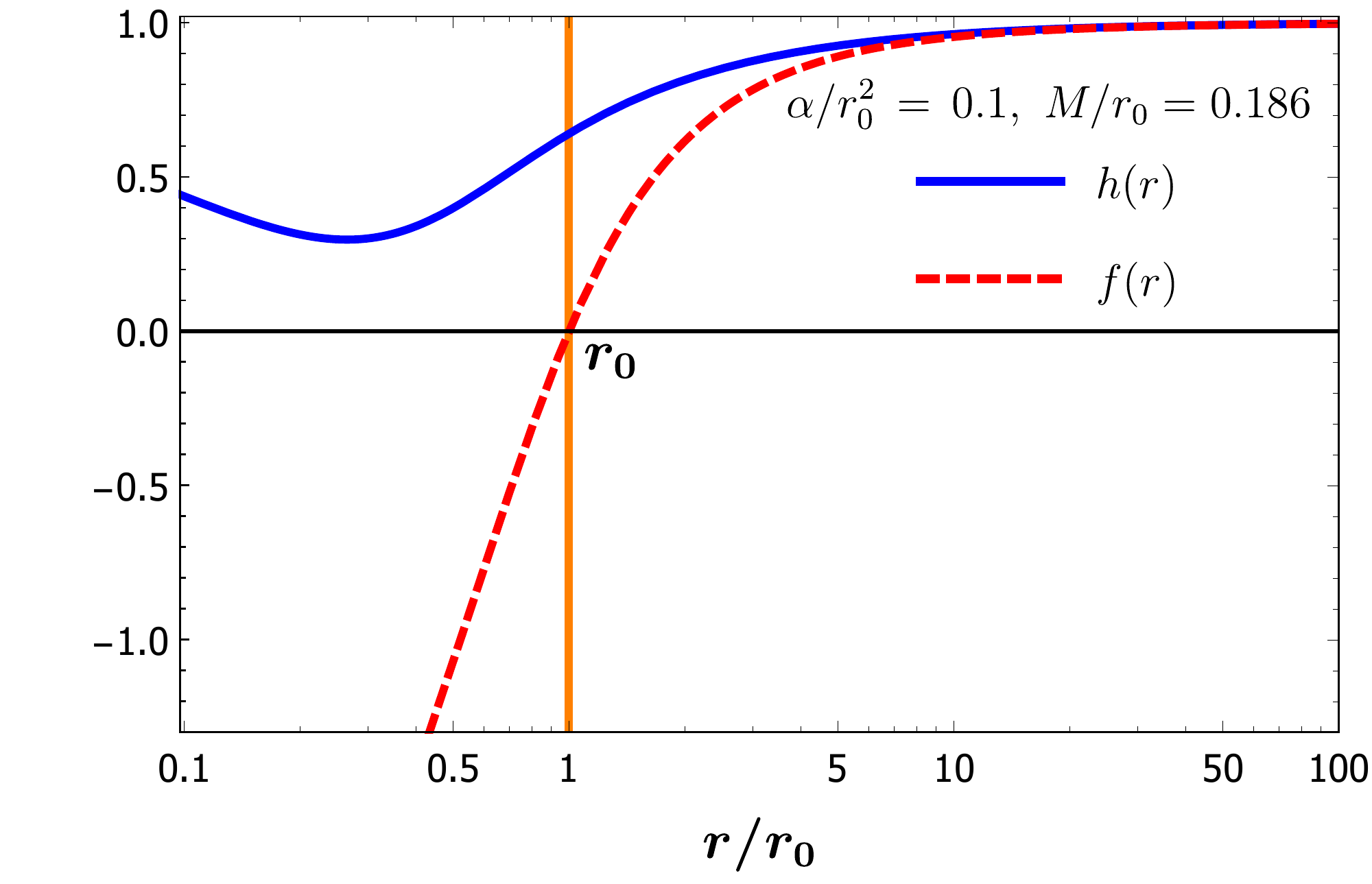}
\\
\hspace*{0.7cm} {(a)} \hspace*{7.5cm} {(b)}  \vspace*{-0.5cm}
\end{center}
\caption{The metric functions $h(r)$ and $f(r)$ for (a) a wormhole solution with $\alpha<M^2$, and (b) a wormhole solution with $\alpha>M^2$.}
  \label{metric-emb1}
\end{figure} 
%%%%%%%%%%%%%%%%%%%%%%%%%%%%%%%%

It is useful to point out the different behaviour of the metric functions in-between the seed Lu-Pang black-hole solution and the wormhole as we do in Fig. \ref{metric-emb1}. For this, we depict the profiles of the metric functions  $h(r)$ and $f(r)$ for two indicative values of $\alpha$. In Fig. \ref{metric-emb1}(a), we take $\alpha<M^2$ where the seed solution is a black hole. The solid (blue) line describes the red-shift function $h(r)$, common to both seed and image solution, which vanishes only at the seed black-hole horizon $r_h$ and asymptotically approaches unity. The dashed (red) line describes the wormhole shape function $f(r)$: the disformal transformation, with $W^{-1}$ given by Eq. \eqref{G}, has changed its profile and forced it to vanish also at the point $r=r_0$. In Fig. \ref{metric-emb1}(b) on the other hand, we depict a solution with $\alpha>M^2$. In this case, the seed is a naked singularity and the metric function $h(r)$ has no roots. The wormhole function $f(r)$ now vanishes only at the location of the throat and again defines, for $r$ in $[r_0, \infty)$ a wormhole metric. 
%Both  cases.
In other words the domain of existence of the wormhole solutions is larger compared to the one of the seed black holes.
In fact, when $\alpha > M^2$ then the minimum throat size occurs at $r_{0}^{min}=\frac{M}{\lambda (2-\lambda)}$. On the other hand the minimum of the redshift function $h$ occurs at $r_1=(\alpha M)^{1/3}$ where $h(r_1)=1-\frac{M^{2/3}}{\alpha^{1/3}}$. From (\ref{alpha_bound}) we then see that $r_{0}^{min}=r_1$ ie., the minimum throat size for $\alpha>M^2$ is that of the minimum of the redshift function h. 

In order to have a geometric picture of the wormhole geometry we can define the circumferential radius $R_c$, 
%%%%%%%%%%
\begin{equation}
    R_c=\frac{1}{2\pi} \int_{0}^{2 \pi} \sqrt{g_{\varphi\varphi}} |_{\theta=\pi/2}\,d\varphi\,.
\end{equation}
If we employ the proper radial distance $\xi$, 
%which for the coordinate system of Eq. \eqref{wormhole} is defined via the relation
%%%%%
\begin{equation}
    d\xi=\pm \,\frac{dr}{\sqrt{f}}\,,
\end{equation}
we may write 
%%%%%%%
\begin{equation}
    \frac{dR_c}{d\xi} = \frac{dR_c}{dr}\,\frac{dr}{d\xi} = \pm \sqrt{f}.
\end{equation}
Therefore, a root $r_0$ of the metric function $f(r)$, or equivalently a root of the disformal transformation function $W^{-1}$, corresponds to an extremum of the
circumferential radius $R_c$. This extremum is a wormhole throat if it is a mimimum \cite{MT}, i.e. if,
%%%%%%%%
\begin{equation}
    \frac{d^2R_c}{d\xi^2}\biggr|_{r_0} =  \frac{d \sqrt{f}}{dr}\,\frac{dr}{d\xi}\biggr|_{r_0} = \frac{f'}{2}>0\,.
\end{equation}
As a result, a wormhole geometry may be indeed realised only if the metric function $f(r)$ satisfies an additional condition, the so-called ``flaring-out'' condition, which in our case takes the form $f'(r_0)>0$. This condition ensures that the spacetime geometry flares out as the radial coordinate increases from the throat towards asymptotic infinity. 

%Note  geometry.

\begin{figure}[t!] 
%\lbfig{Fig_phi} 
\begin{center}
\hspace{0.2cm} \hspace{-0.5cm}
\includegraphics[height=.23\textheight, angle =0]{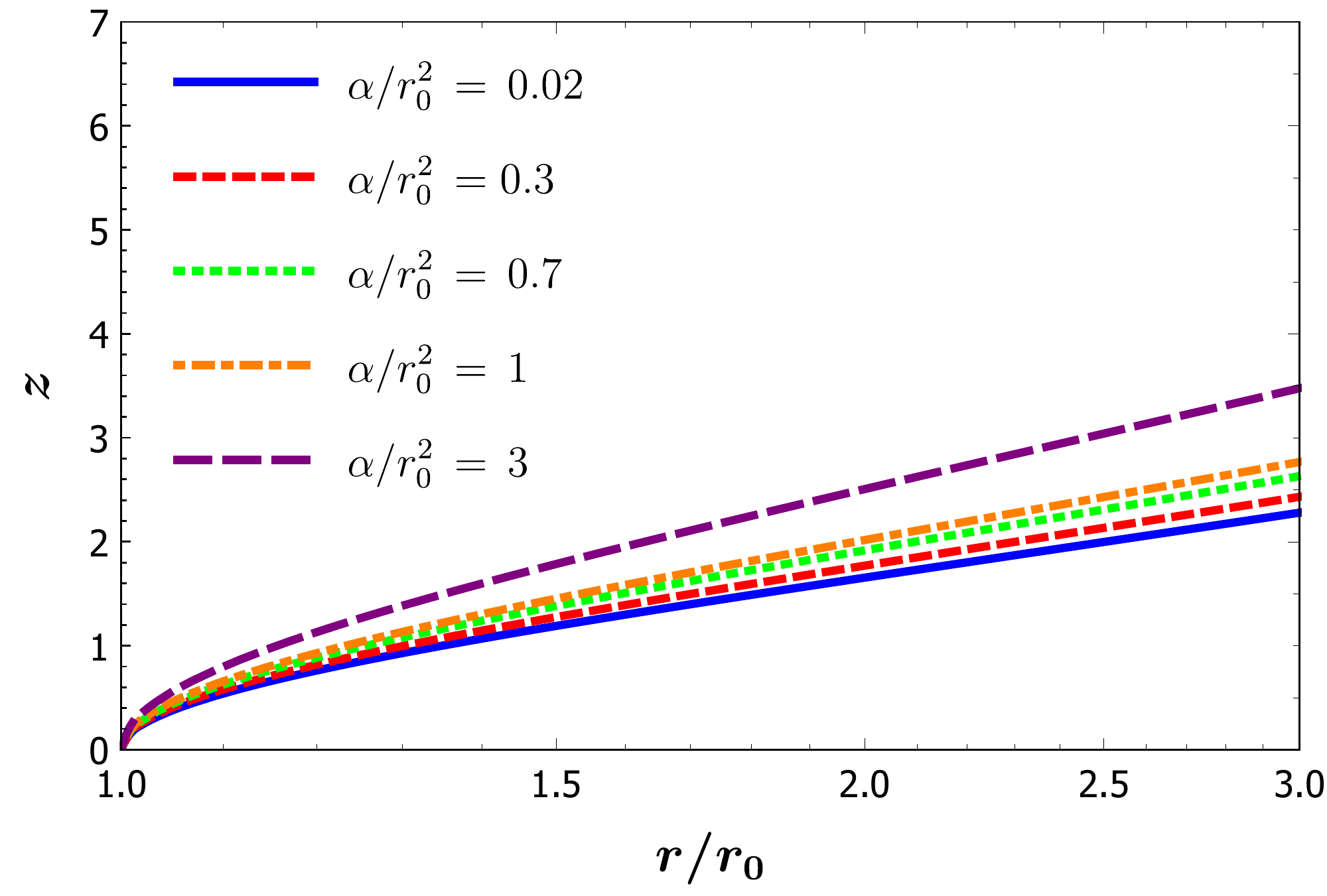}
 \hspace{1cm} \hspace{-1.2cm} 
\includegraphics[height=.2\textheight, angle =0]{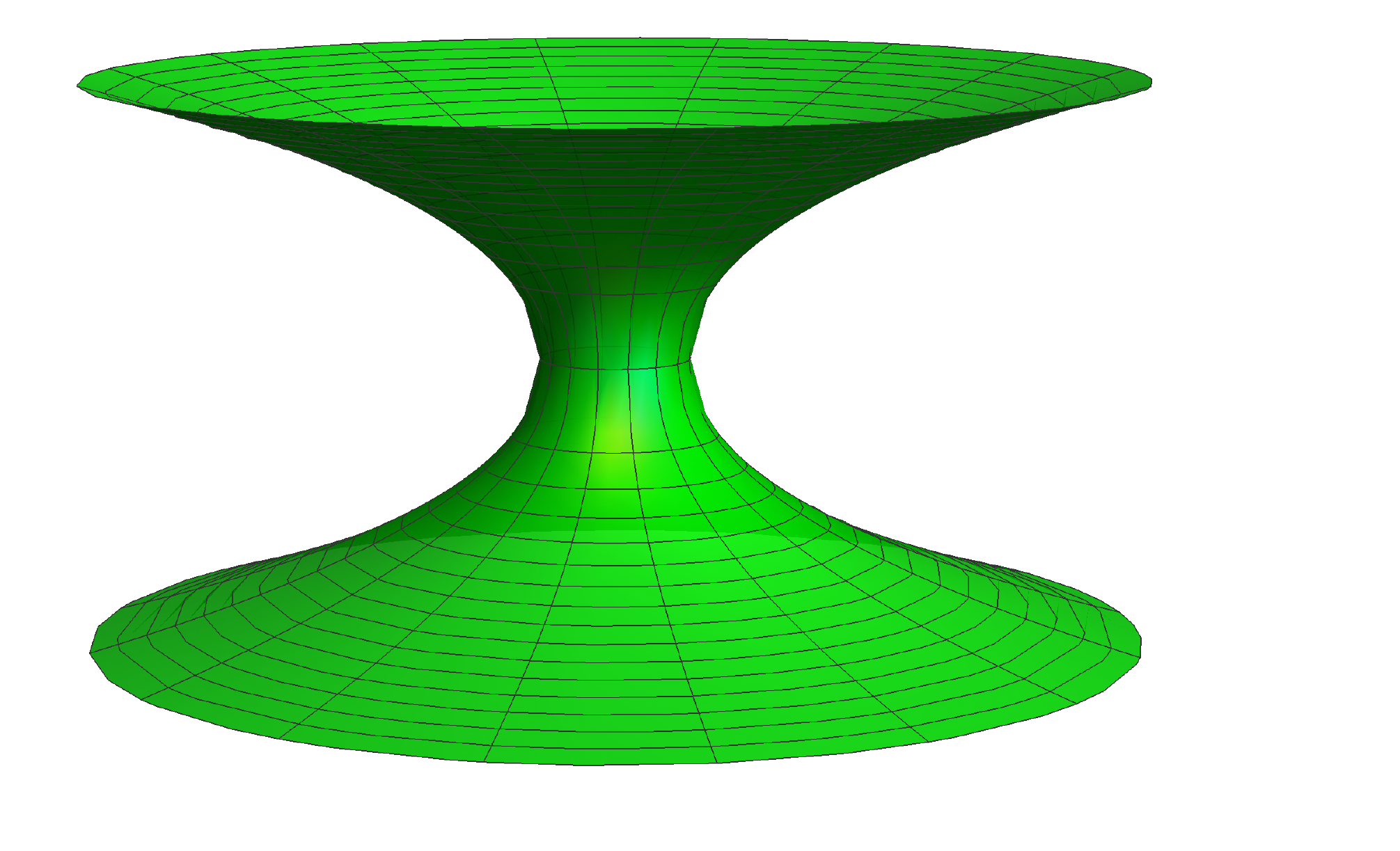}
\\
\hspace*{0.7cm} {(a)} \hspace*{7.5cm} {(b)}  \vspace*{-0.5cm}
\end{center}
\caption{(a) The embedding diagram for five wormhole solutions, and (b) the 3-D representation  of the wormhole solution with  $\alpha/r_0^2=0.02.$}
  \label{emb1}
\end{figure}

A useful way to visualize the wormhole geometry is as a surface embedded in three dimensional flat space. Given our symmetries we consider $t=const$ and $\theta=\pi/2$. The line-element then reads,
\begin{equation}
    ds^2=  \frac{dr^2}{f}+r^2 d\varphi^2\label{eqp}.
\end{equation}
In order to find the isometric embedding, we consider a three-dimensional flat, Euclidean hypersurface and demand that its line-element coincides with Eq.  (\ref{eqp}) --  then, the two manifolds have the same geometry. This yields
\begin{equation}
    \frac{dr^2}{f}+r^2 d\varphi^2=dz^2+d\rho^2+\rho^2d\varphi^2,
\end{equation}
where $z$, $\rho$, and $\varphi$ are a set of cylindrical coordinates on the hypersurface. Considering $z$ and $\rho$ as functions of $r$ and matching the coefficients of $d\varphi^2$ and $dr^2$, we easily find $\rho=r$ as well as 
\begin{equation}
    \frac{dz}{dr} = \pm \sqrt{\frac{1}{f(r)} -1}\,, \label{slope}
\end{equation}
or, equivalently,
\begin{align}
   z(r)=\pm \int_{r_0}^r \sqrt{\frac{1}{f(r')} -1}\,dr'. \label{zeq}
\end{align} 
Therefore $\{\rho(r), z(r)\}$ is a parametric representation of a slice of the embedded $\theta = \pi/2$-plane for a fixed value of the $\varphi$ coordinate. Using the analytic form of the shape function $f(r)$, given by Eqs. \eqref{fun-dis} and \eqref{G}, in Fig. \ref{emb1}(a), we plot the embedding function $z(r)$ for five wormhole solutions parametrized by different values of $\alpha/r_0^2$ and $\lambda=0.2$. A 3-D representation of the wormhole's geometry is given by the embedding surface, which follows by considering the revolution of $z(r)$ over a $2\pi$ $\varphi$-angle around the $z$ axis. This surface  is presented in  Fig. \ref{emb1}(b) for the indicative case with $\alpha/r_0^2=0.02$.

%%%%%%%%%%%%%%%%%%%%%%%%%%
%%%%%%%%%%%%%%%%%%%%%%%%%%%%%%%%

We observe that the embedding surface exhibits the desired `flaring-out' described earlier. Its slope reflects the magnitude of curvature at a given point of the spacetime: it becomes horizontal at the asymptotically-flat regime, where the curvature is zero, and vertical at the throat, where the curvature takes its highest, albeit finite\,\footnote{In terms of the proper distance, which is the correct coordinate-invariant quantity, the slope of the embedding surface is given by:  $dz/d\xi = \sqrt{1-f(r)}$. This quantity goes smoothly to zero at asymptotic infinity, where $f \rightarrow 1$, and to unity near the throat, where $f$ vanishes; therefore, it remains always finite.}, value. If we invert the function $z(r)$ to obtain $r(z)$, it is clear that the presence of the throat corresponds to the existence of a minimum for the latter function. Indeed, from Eq. \eqref{slope}, we immediately obtain that the first derivative 
%%%%%%%%%
\begin{equation}
    \frac{dr}{dz} = \pm \sqrt{\frac{f(r)}{1-f(r)}}
    \label{slope_min}
\end{equation}
%%%%%%%%%%%
vanishes at the throat since there $f \rightarrow 0$. For the second derivative, we readily obtain
%%%%%%%
\begin{equation}
  \frac{d^2r}{dz^2} =  \frac{f'(r)}{2\,(1-f)^2}\,.
\end{equation}
Therefore, in order to have a minimum for the function $r(z)$, and thus a throat, we should have $f'(r_0)>0$. This is again the `flaring-out' condition that the shape function needs to satisfy in order to realise the wormhole geometry.

\section{Regular and traversable wormholes}
\setcounter{equation}{0}

%%%%%%%%%%%%%%%%%%%%%%
%
\begin{figure}[t] 
%\lbfig{Fig_phi} 
\begin{center}
\hspace{0.0cm} \hspace{-0.5cm}
\includegraphics[height=.21\textheight, angle =0]{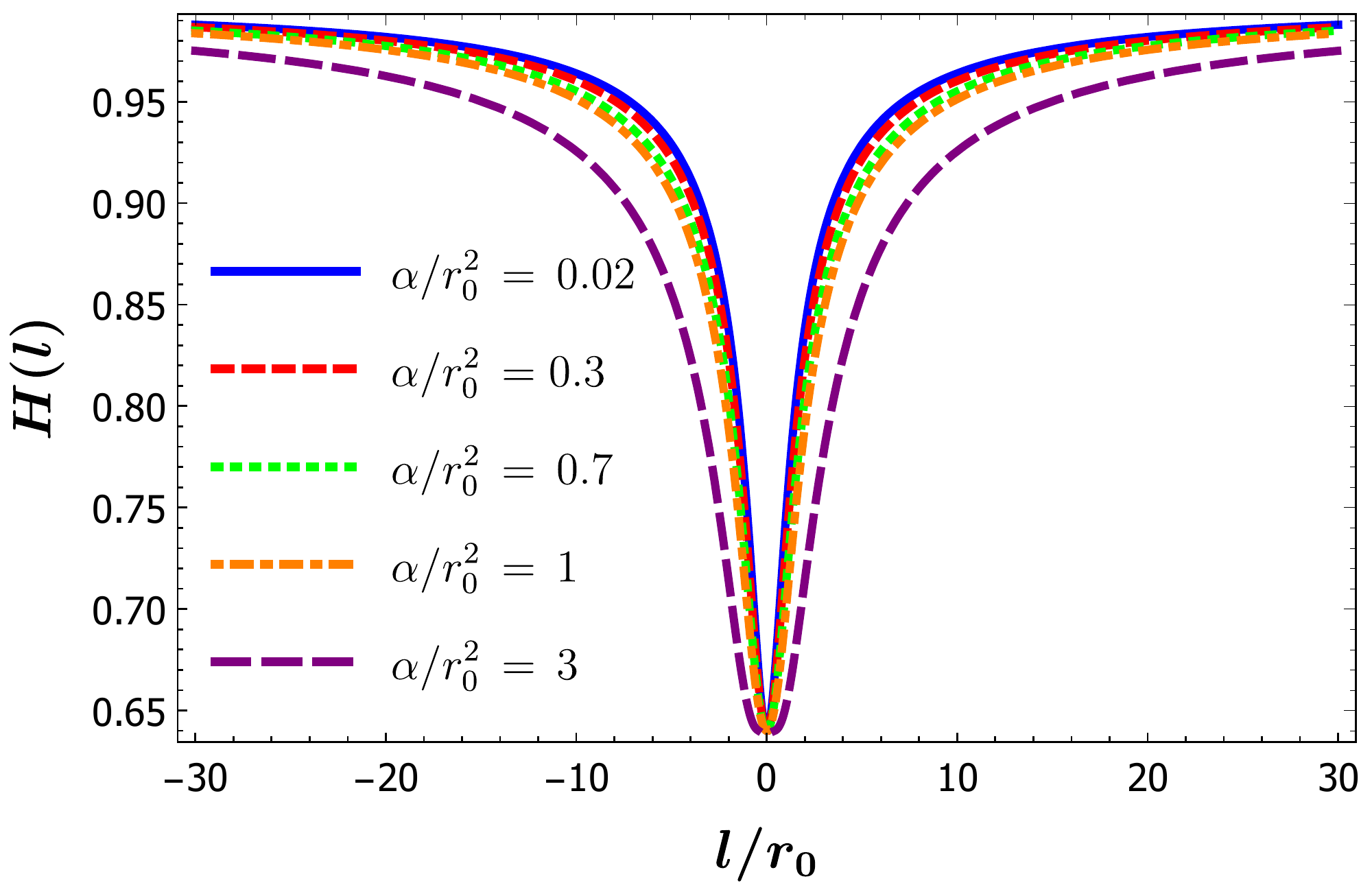}
\hspace{0.7cm} \hspace{-0.5cm}
\includegraphics[height=.21\textheight, angle =0]{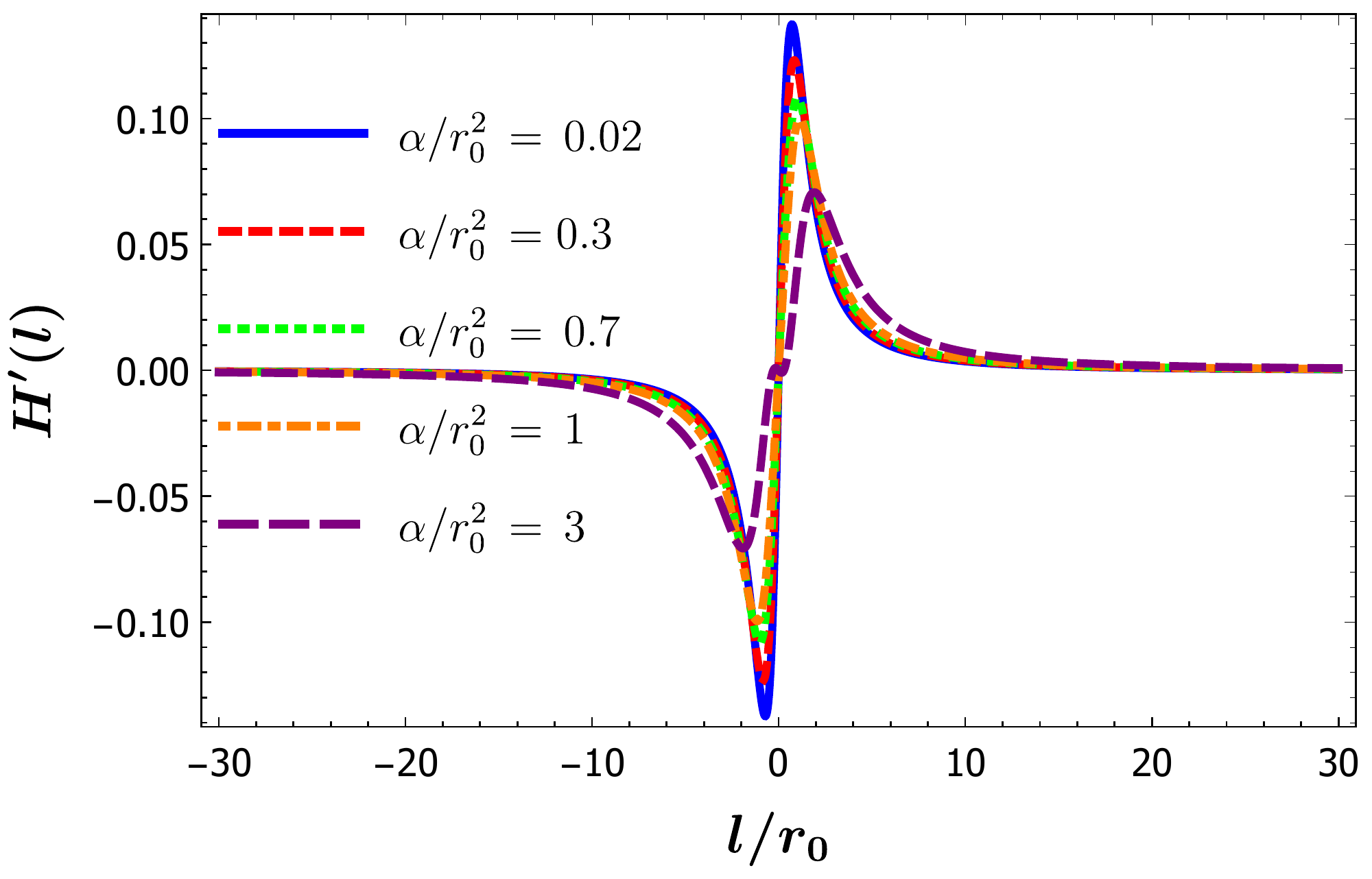}
\\
\hspace*{0.7cm} {(a)} \hspace*{7.5cm} {(b)}  \vspace*{-0.1cm}
\hspace{0.2cm} \hspace{-0.5cm}
\includegraphics[height=.21\textheight, angle =0]{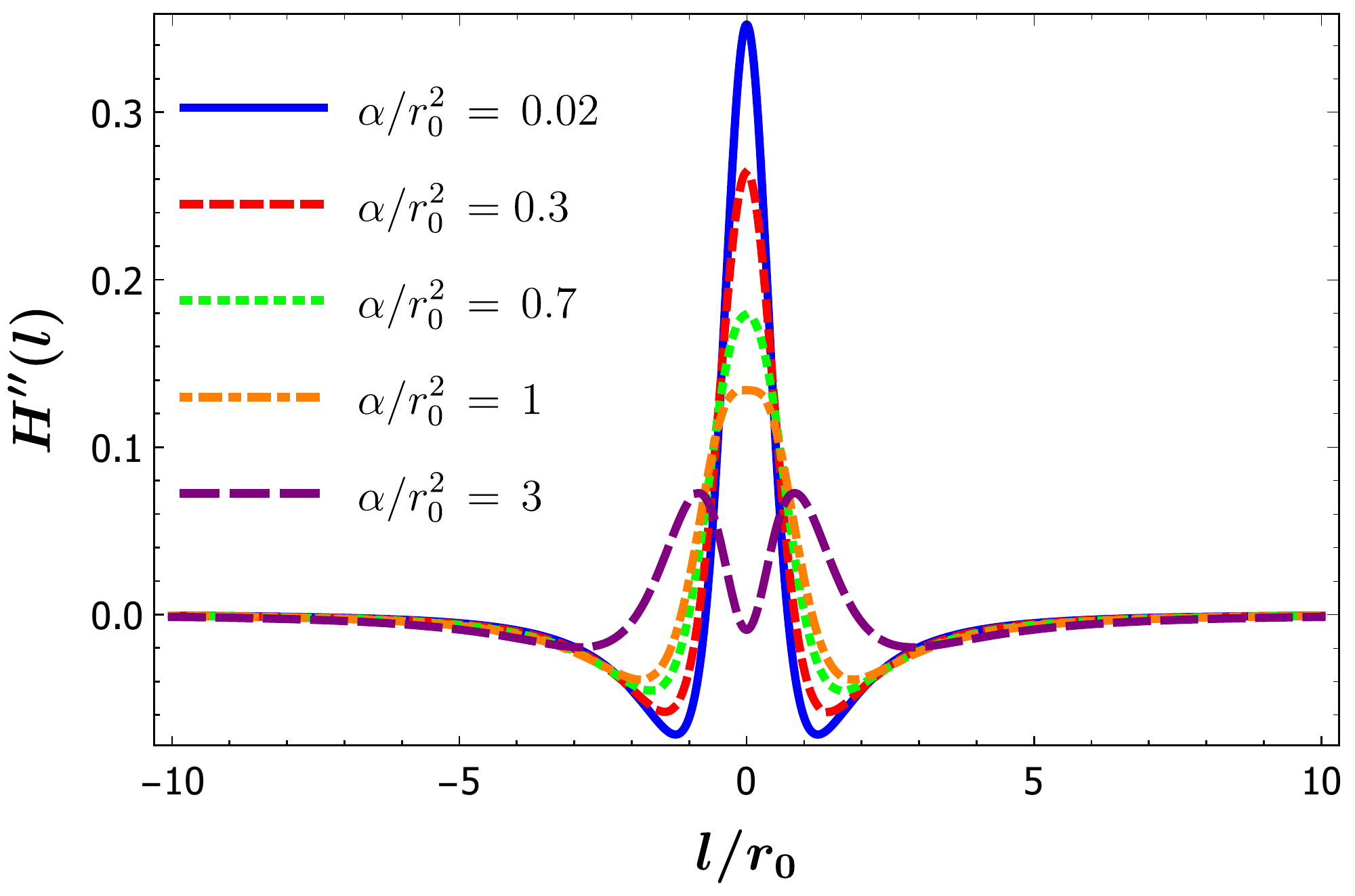}
\\[1mm]
\hspace*{0.7cm} {(c)}  \vspace*{-0.5cm}
\end{center}
\caption{The  metric function $H(l)$,  its  first and second derivative in terms of the radial coordinate $l$, for five different solutions parametrised by the value of $\alpha/r_0^2$.}
  \label{hh1}
\end{figure} 
%%%%%%%%%%%%%%%%%%%

\begin{figure}[t] 
%\lbfig{Fig_phi} 
\begin{center}
\hspace{0.0cm} \hspace{-0.5cm}
\includegraphics[height=.21\textheight, angle =0]{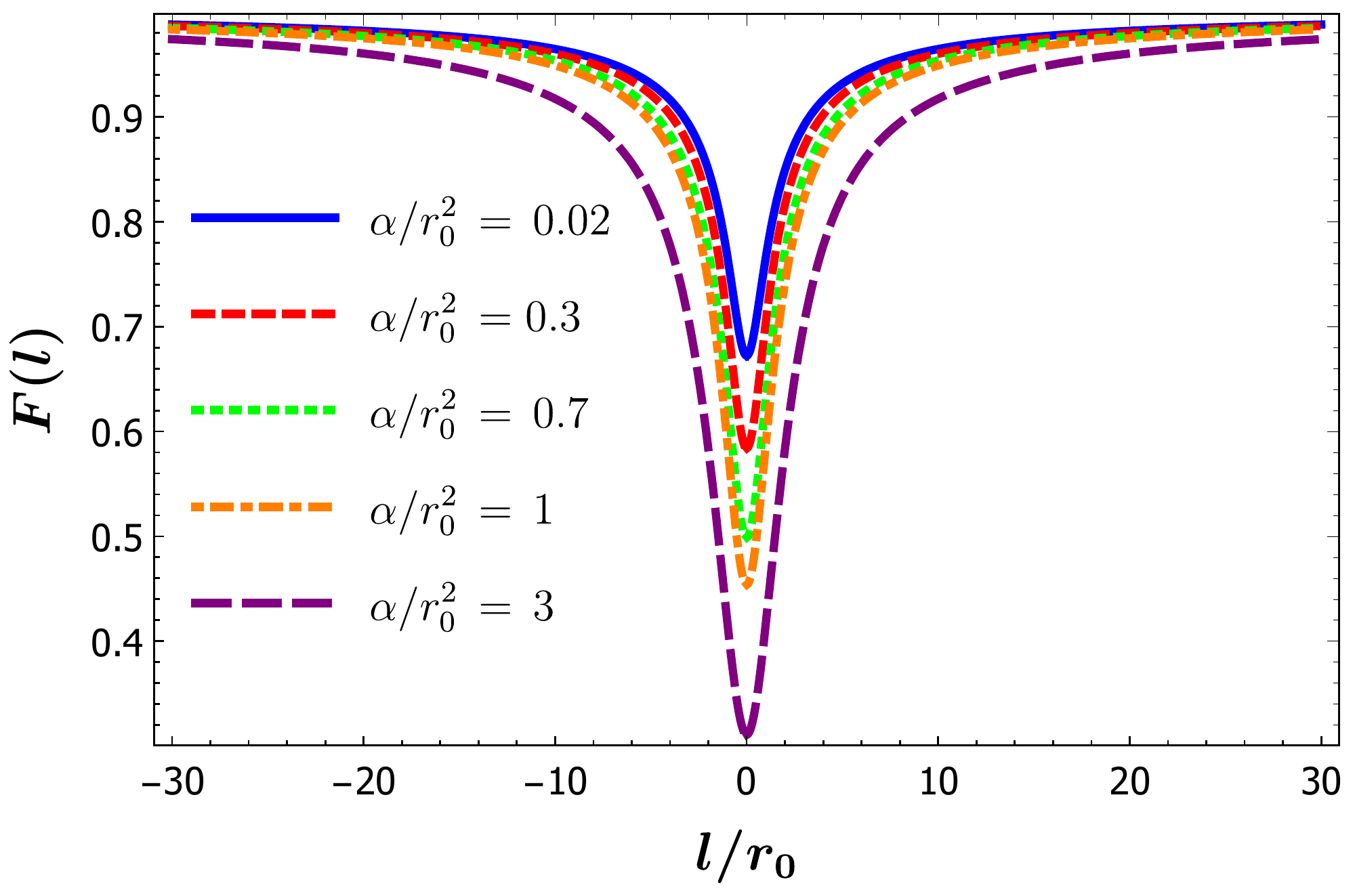}
\hspace{0.7cm} \hspace{-0.5cm}
\includegraphics[height=.21\textheight, angle =0]{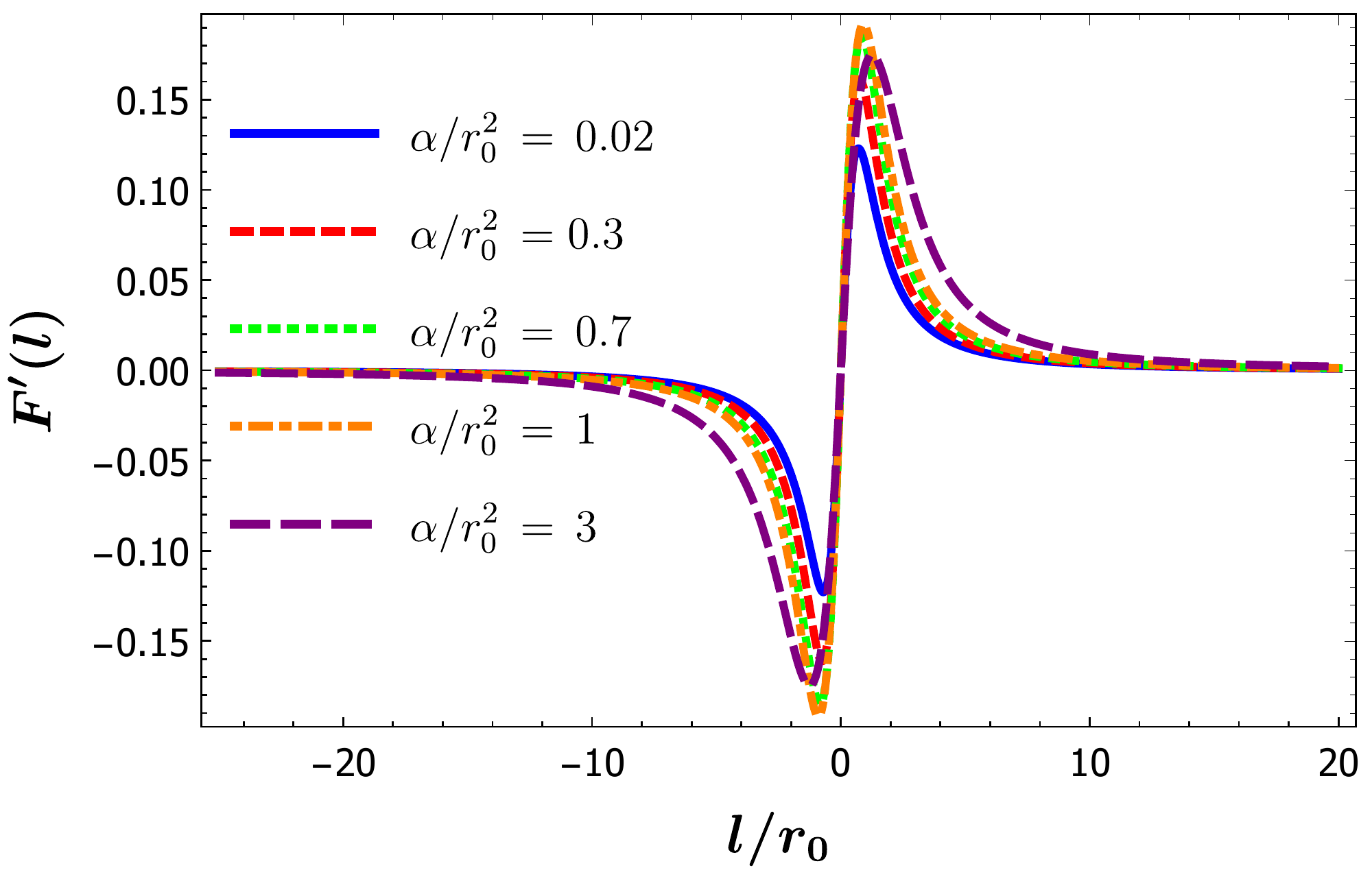}
\\
\hspace*{0.7cm} {(a)} \hspace*{7.5cm} {(b)}  \vspace*{-0.1cm}
\hspace{0.2cm} \hspace{-0.5cm}
\hspace{0.7cm} \hspace{-0.5cm}
\includegraphics[height=.21\textheight, angle =0]{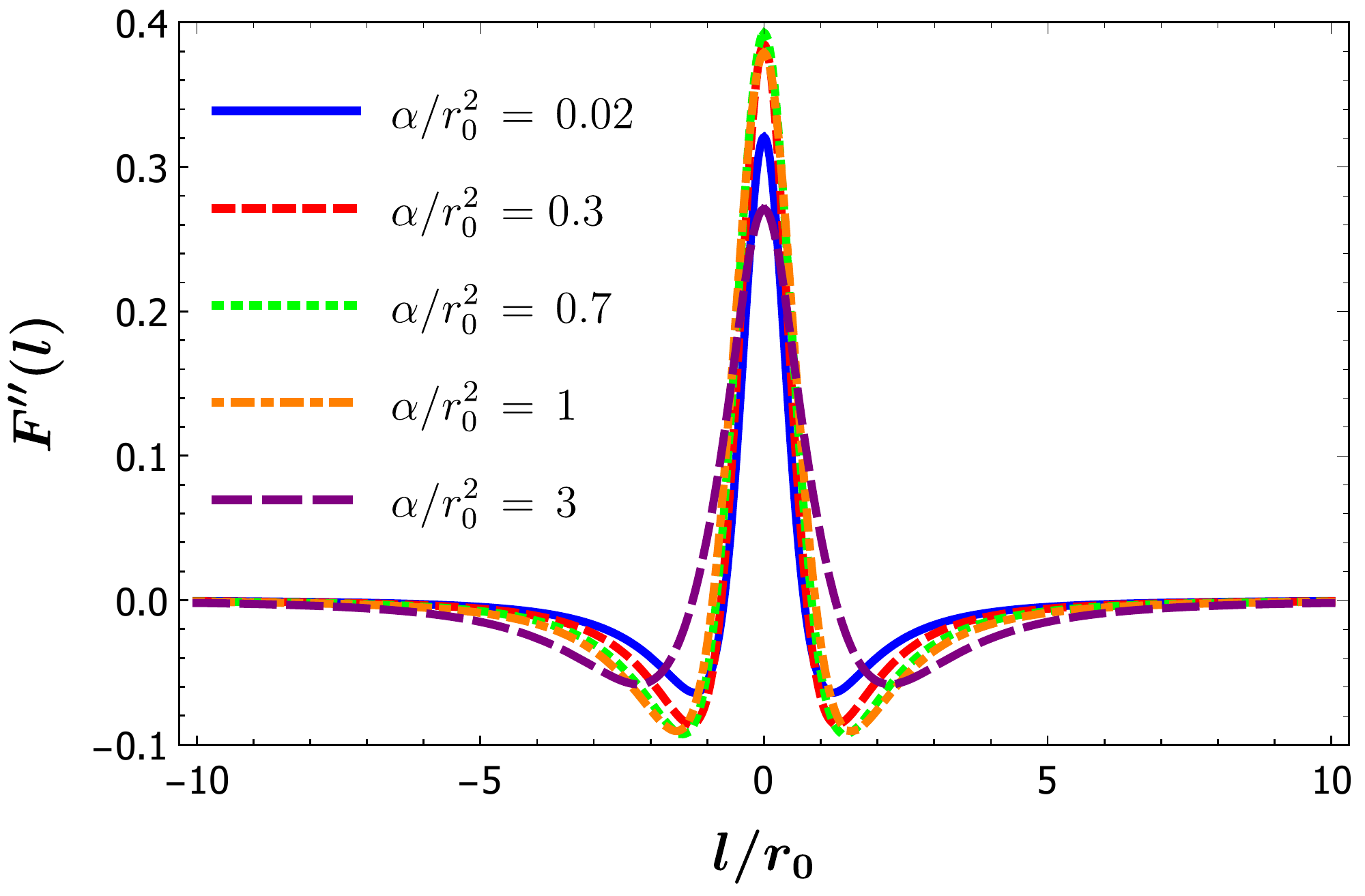}
\\[1mm]
\hspace*{1.0cm} {(c)}  \vspace*{-0.5cm}
\end{center}
\caption{The  metric function $F(l)$, its  first and second derivative in terms of the radial coordinate $l$, for the same set of solutions parametrised by
$\alpha/r_0^2$.}
  \label{Ffs}
\end{figure}

The physically relevant part of the wormhole spacetime is the one described by $r \geq r_0$, where $f(r) \geq 0$, depicted in the embedding diagrams of Fig. \ref{emb1}(a). The wormhole geometry possesses a throat connecting two asymptotically-flat regions, as depicted in Fig. \ref{emb1}(b) where $\partial_t$ is a global Killing vector. At $r=r_0$ our coordinate system breaks down and we need to determine if this is poor choice of coordinates or an unavoidable curvature singularity. Consider the coordinate transformation  $r^2=l^2+r_0^2$, or $l=\pm \sqrt{r^2-r_0^2}$; therefore, as $r \in [r_0, \infty)$ we have that  $l\in (-\infty,+\infty)$. The new coordinate discards the region $r<r_0$ and double covers the region $r\geq r_0$ thus accommodating two asymptotically-flat regions. In terms of the new coordinate $l$, the throat is located at $l=0$ and the line-element \eqref{wormhole} takes the following form:  
%%%
\begin{equation}
    ds^2=-H(l) \,dt^2+\frac{dl^2}{F(l)}+(l^2+r_0^2)\,d\Omega^2,
\end{equation}
%%%%%%%
where
%%%%%
%
\begin{equation}
\label{worml}
  H(l)=h(r(l)), \quad \text{and} \quad  F(l)=\frac{f(r(l))\,(l^2+r_0^2)}{l^2}.
\end{equation}
%
%%%%%
 As we observe in Figs. \ref{hh1} and \ref{Ffs}, the metric functions $H(l)$ and $F(l)$ are finite and $C^2$. Indeed, it turns out that in these coordinates the wormhole metric is $C^2$ regular for all $l \in (-\infty, +\infty)$, including of course the throat location at $l=0$. 
%In Fig. \ref{hh1}. 
Note that, after the coordinate change, the metric function $F(l)$ does not vanish at $l=0$ but instead assumes a constant value.  From the profiles of both metric functions, we observe that as the coupling parameter of the theory $\alpha/r_0^2$ increases, the slope of the corresponding curves at the asymptotically flat regimes increases too: this signifies that the wormhole becomes heavier but also larger.
%- we solution. 
{\footnote{Figures \ref{hh1} and \ref{Ffs} have been constructed for the indicative case of $\lambda=0.2$, however, the depicted behaviour does not qualitatively change as the value of $\lambda$ varies in the range $(0,1)$.}}

%%%%%%%%%%%%%%%%%%%%%%%%%%%%
%
%%%%%%%%%%%%%%%%%%%%%%%%%%%

According to Figs. \ref{hh1} and \ref{Ffs}, the metric functions $H(l)$ and $F(l)$ as well as their second derivatives are symmetric under the change $l \rightarrow -l$, and assume nonvanishing values at $l=0$. In contrast, their first derivatives are antisymmetic under the same change and vanish at the throat. This is due to the fact that, close to the throat,  both metric functions may be written  as power-law expansions in terms of the coordinate $l$ which contain only even powers, namely
%%%%%%
\be
H(l)= h_0 + h_2\,l^2 + O(l^4)  \,, \qquad F(l) = f_0 + f_2\,l^2 + O(l^4) ,
\ee
%%%%%%
where $(h_i, f_i)$ are functions of the parameters of the theory given in (\ref{values1}-\ref{values4}).
Due to the vanishing of $H'(l)$ and $F'(l)$ at $l=0$, both metric functions reach their minimum values at $l=0$ and both increase towards the two asymptotic infinities.  This behaviour allows for a smooth transition from positive to negative $l$ values, and no discontinuities or cusp points appear in the metric function derivatives. An important implication of this is that {\it no additional distribution of matter} needs to be introduced around the wormhole throat for the smooth patching of the two $l$-regimes.

%%%%%%%%%%%%%%%%%%%%%%%%%%%%%%%%%
\begin{figure}[t!] 
%\lbfig{Fig_phi} 
\begin{center}
\hspace{0.0cm} \hspace{-0.5cm}
\includegraphics[height=.19\textheight, angle =0]{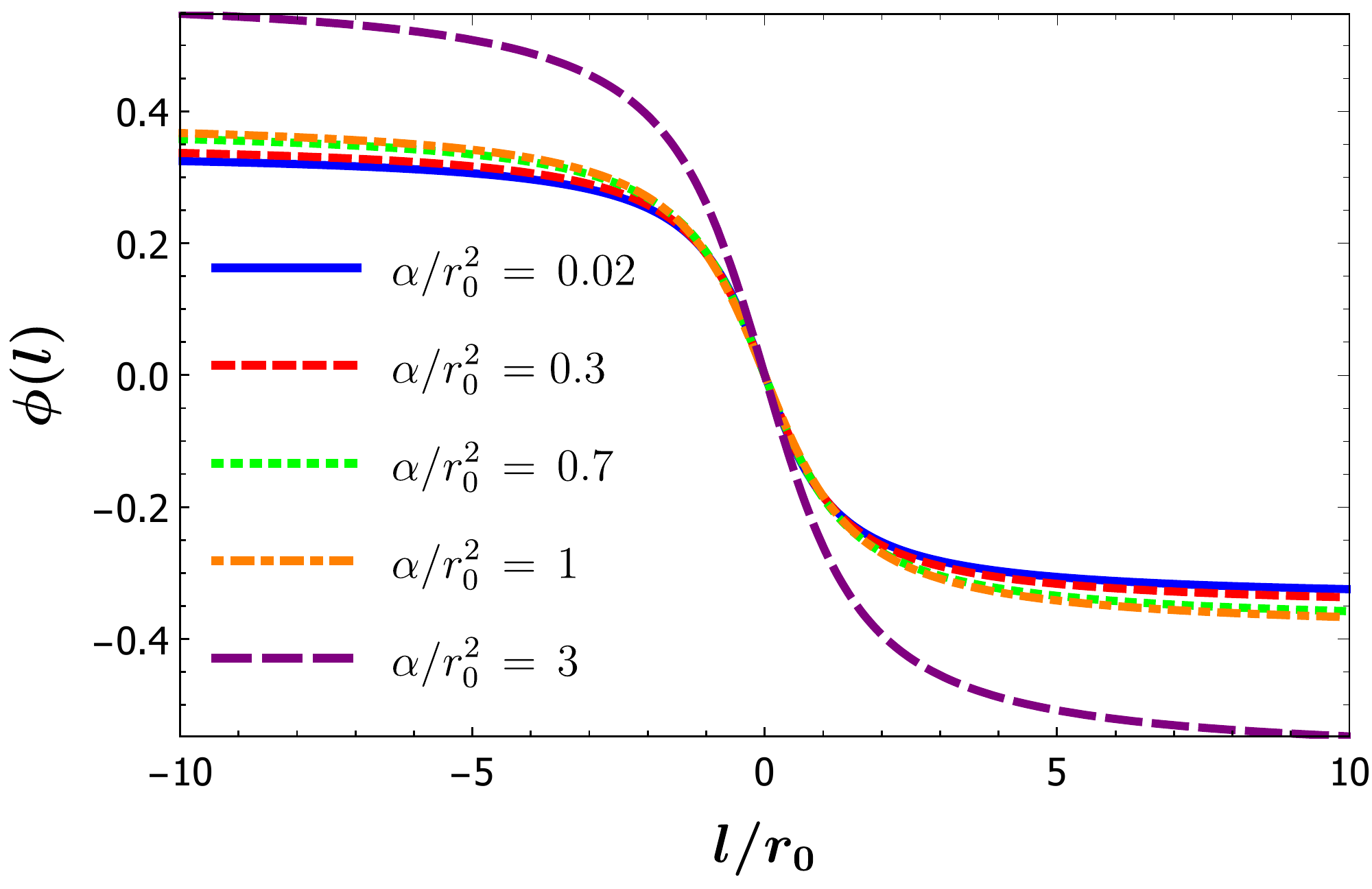}
\hspace{0.7cm} \hspace{-0.5cm}
\includegraphics[height=.19\textheight, angle =0]{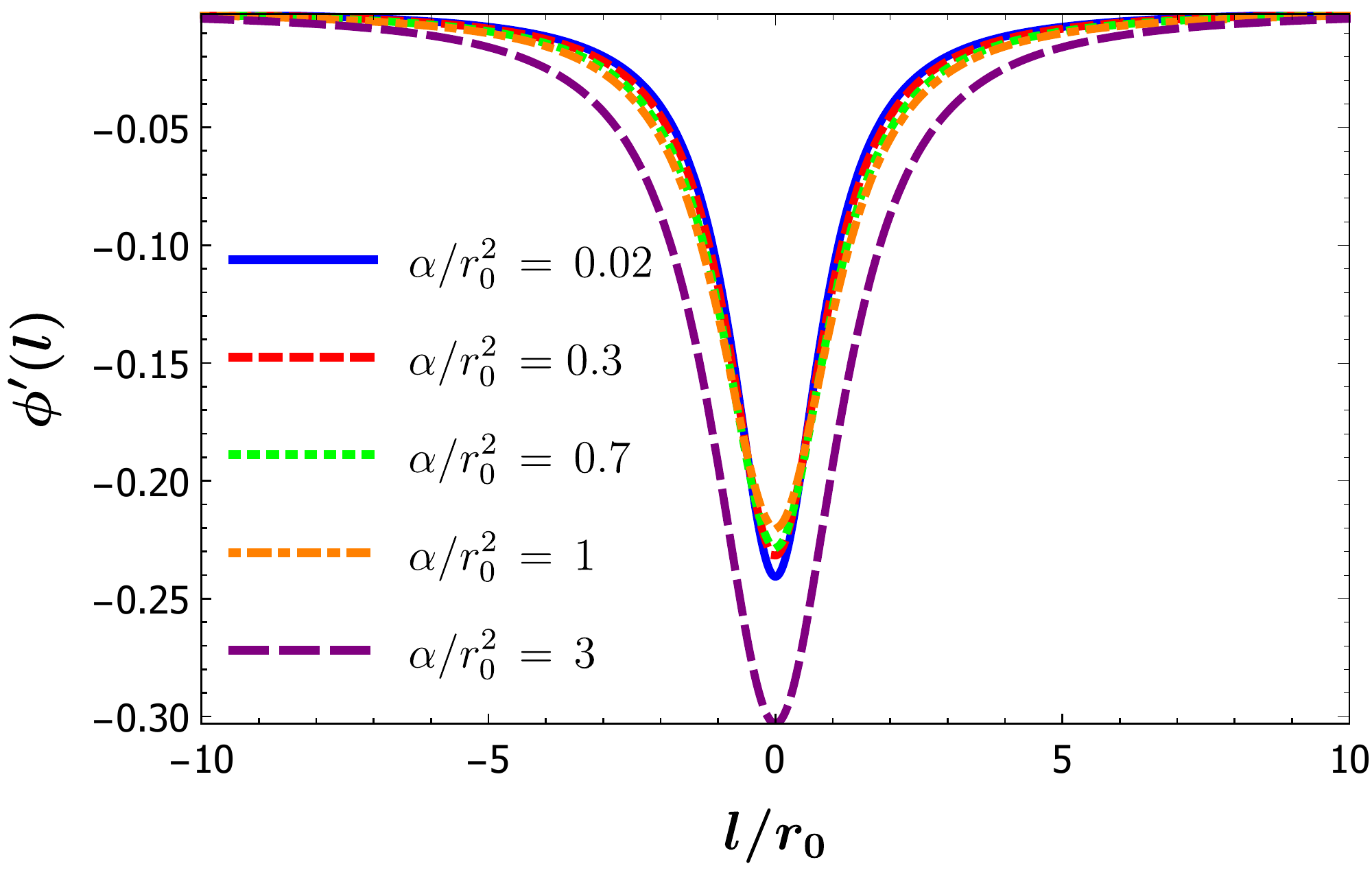}
\\
\hspace*{0.7cm} {(a)} \hspace*{7.5cm} {(b)}  \vspace*{-0.1cm}
\hspace{0.2cm} \hspace{-0.5cm}
\hspace{0.2cm} \hspace{-0.5cm}
\includegraphics[height=.19\textheight, angle =0]{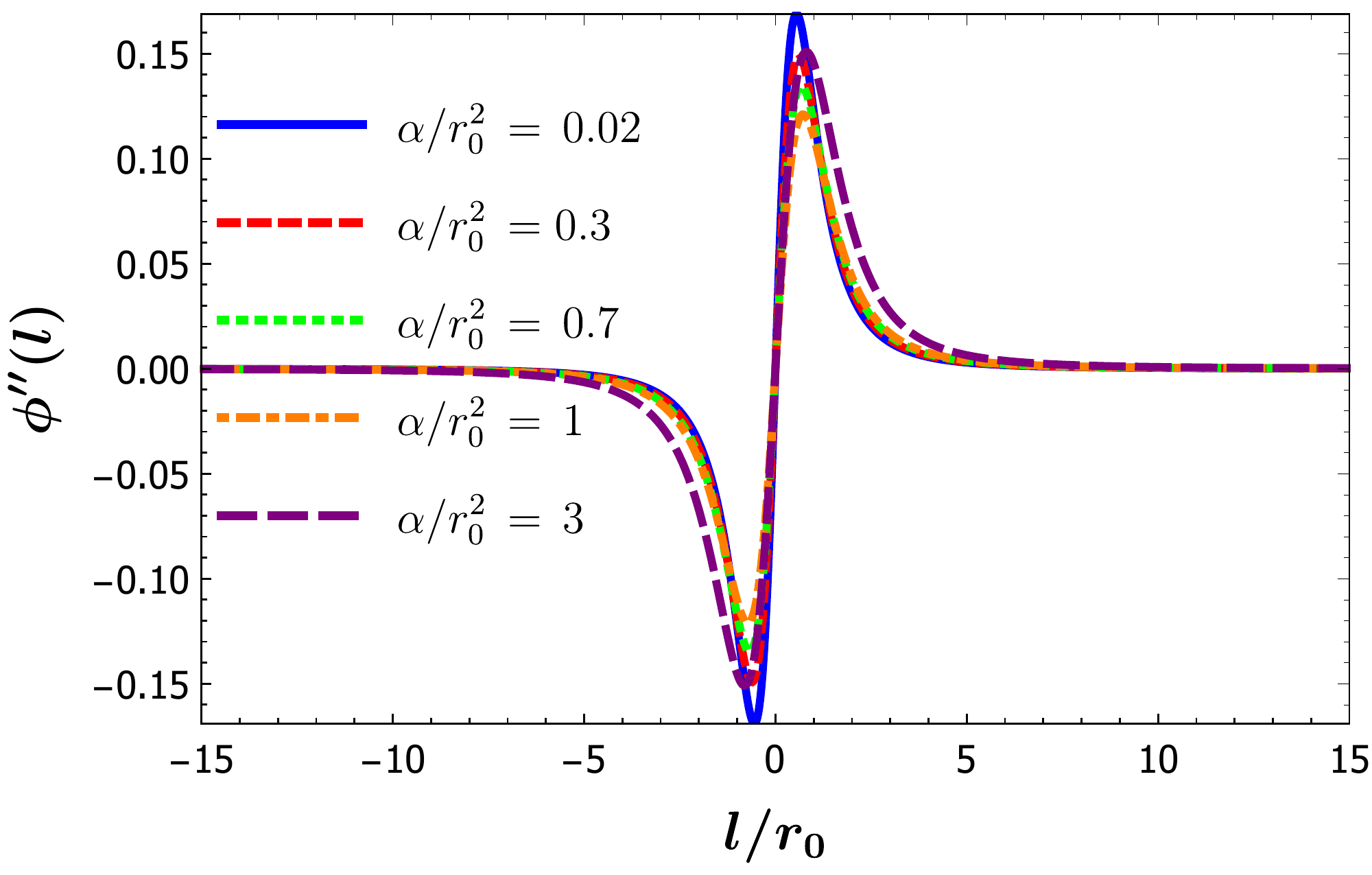}
\\[1mm]
\hspace*{0.8cm} {(c)}  \vspace*{-0.5cm}
\end{center}
\caption{The scalar field $\phi(l)$,  its  first and second derivative in terms of the radial coordinate $l$, for the same set of solutions parametrised by
$\alpha/r_0^2$.}
  \label{phips}
\end{figure} 
%%%%%%%%%%%%%%%%%%%%%%%%%

Finally we turn to the profile of the scalar field, the first derivative of which is also analytically known and given in Eq. (\ref{phih}). In order to determine $\phi$ itself, we first take the expansion of $\phi'(l)$ at $l=0$ and then integrate with respect to $l$ to find
%%%%%%
\begin{equation}
    \phi(l)=\f_0+\phi_1 \,l+\mathcal{O}(l^3)\,, \label{phi_expansion}
    \end{equation}
%%%%%%%%%
where 
%$$   }  },$$
$$\phi_1=\frac{\lambda}{r_0(\lambda-1)},$$
and given the shift symmetry we may choose $\phi_0=0$.
Therefore, at the wormhole throat, the scalar field vanishes whereas its first derivative is non zero. %Then, lar  $\phi(0)=0$. 
In Fig. \ref{phips}, we depict the scalar field $\phi(l)$ and its  derivatives -- for consistency, we use the same five solutions employed in the construction of Figs.  \ref{hh1} and \ref{Ffs}. We observe that all three quantities are everywhere regular with the scalar field $\phi$ and its second derivative being antisymmetric under the change $l \rightarrow -l$, and thus vanishing at $l=0$, whereas its first derivative is symmetric in agreement with Eq. (\ref{phi_expansion}). The scalar field approaches a constant value at the two asymptotic infinities where both of its derivatives vanish. The transition from the positive to the negative $l$-regime is again smooth and the $\phi$ profile is continuous at $l=0$. We would like to note that Figs.  \ref{phips}(b,c) depict the numerical values of the first and second derivative of the scalar field, however,  we have verified that these are in excellent agreement with the analytic expressions for $\phi'$ and $\phi''$ following from Eq. (\ref{phih}). 

In conclusion, the wormhole solution \eqref{worml} is regular at the throat and throughout spacetime. Due to its $C^2$ regularity at the throat, it is also free of any additional local (or non local sources of matter). As we will demonstrate in the next section, where the properties of the solution are studied in detail, our wormhole is also traversable by light and matter. In fact, a potential traveler will experience small tidal forces and acceleration effects upon choosing particular values of the parameters of the theory.

%%%%%%%%%%%%%%%%%%%%%%%%%%%%%%%%%%%%%%%%%%%%%%%%%%%%%%%%%%%%%%%%%%%%%%%%%%%%%%%%%%%%%%%

\section{Wormhole properties}
\setcounter{equation}{0}

Let us first make a connection to the classical wormhole analysis of Morris and Thorne \cite{MT} as seen from the perspective of General Relativity. In order to do this we can define an ${\it{effective}}$ energy-momentum tensor such that our field equations take the suggestive form
\begin{equation}
\label{eff}
G_{\mu \nu}=T_{\mu \nu}^{\mbox{\tiny{eff}}}\,. 
\end{equation}
Essentially $T_{\mu \nu}^{\mbox{\tiny{eff}}}$ entails all the effects coming from the beyond Horndeski theory apart from the GR Einstein term (which amounts to the constant term in $G_4$). We emphasize that $T_{\mu \nu}^{\mbox{\tiny{eff}}}$ is not ordinary matter but encompasses here collectively the effects of modified gravity.

According to \cite{MT}, the energy conditions of this effective energy-momentum tensor have to be violated near the throat. This is in fact the outcome of the flaring-out condition, which as we showed imposes a certain behaviour on the shape function $f(r)$ and, through the effective Einstein's equations, on the matter content of the theory. Let us see how this applies in the case of our solutions. We will focus on the Null Energy Condition (NEC)  \cite{Harko:2013yb} \cite{Kanti:2011jz} \cite{Antoniou:2019awm},
\begin{align}
T_{\mu\nu}^{\mbox{\tiny{eff}}}\,n^\mu n^\nu \ge 0,
\end{align}
where $n^\mu$ is any null vector satisfying $n^\mu n_\mu=0$. Choosing respectively $n^\mu=(1, \sqrt{-g_{tt}/g_{rr}},0,0)$ and  $n^\mu=(1, 0, \sqrt{-g_{tt}/g_{\theta\theta}},0)$ while making use of (\ref{eff}) we have that the NEC gives,
%the  the Einstein tensor, namely 
%
\begin{align}
& G_r^r-G_t^t \ge 0, \\[2mm]
& G_\theta^\theta -G_t^t \ge 0.
\end{align}
Employing the form of the line-element \eqref{wormhole}, we find, 
%%%%
\begin{align}
G_t^t &= -\frac{1}{r^2} +\frac{h}{r^2 W} -\frac{h W'}{W^2 r} +\frac{h'}{r W}\,, \label{Gtt}\\[2mm]
%G_r^r &= -\frac{1}{r^2} +\frac{h}{r^2 W} +\frac{h'}{r W}\,, \label{Grr}\\[2mm]
%G_\theta^\theta &=-\frac{h W'}{2 r W^2} +\frac{h'}{r W} -\frac{W' h'}{4 W^2}+\frac{h''}{2 W} \label{Gthth}\,.
G_r^r-G_t^t &= \frac{h W'}{W^2 r}, \label{Grr}\\[2mm]
G_\theta^\theta - G_t^t &=-\frac{W' h'}{4 W^2}+\frac{h W'}{2 W^2 r}+\frac{h''}{2 W}-\frac{h}{W
   r^2}+\frac{1}{r^2}.
\end{align}
%%%%%%
%where  t the throat $r_0$.
Using that
%%%%%%%
\begin{equation}
   f' = \frac{h'}{W} -\frac{h W'}{W^2}\,, \label{fprime}
\end{equation}
and the fact that $h$ and all its derivatives take constant values at the throat where $W$ vanishes, we find that
%%%%%%%%%%
\begin{align}
(G_r^r-G_t^t) \Bigr|_{r_0}&= -\frac{f'(r_0)}{r_0} <0 , \\[2mm]
(G_\theta^\theta - G_t^t) \Bigr|_{r_0} &= \frac{1}{r_0^2} + \frac{W'}{2W^2}\left(\frac{h}{r}-\frac{h'}{2}\right)\biggr|_{r_0}\,. \label{NEC2r0}
\end{align}
%%%%%%%%%%%
Therefore, if the metric function $f$ satisfies the flaring-out condition $f'(r_0) > 0$, the first NEC is violated at the throat. The second combination on the other hand is not sign-definite but depends on the profile of the red-shift function $h(r)$:  if the right-hand-side of Eq. \eqref{NEC2r0} gives a positive number at the throat, then the second NEC is not violated.

The Weak Energy Condition (WEC) is the NEC supplemented by the positivity of the effective energy density supplied by our modified gravity theory, $\rho_{\mbox{\tiny{eff}}} \geq 0$. We find,
%%%%%%%%%
\begin{equation}
    \rho_{\mbox{\tiny{eff}}}=-G^t_t=
    \frac{1}{r^2} -\frac{h}{r^2 W} +\frac{h W'}{W^2 r} -\frac{h'}{r W}\,.
\end{equation}
Thus, at the throat where $W^{-1}=0$ and $h$ is finite,
%%%%%%%%
\begin{equation}
    \rho_{\mbox{\tiny{eff}}}|_{r_0}= \frac{1}{r_0^2}  -\frac{f'(r_0)}{r_0} \,,
    % =\frac{R(r_0)}{2}-\frac{W' h'}{2 W^2}\|_{r_0}\,.
\end{equation}
where we have also used Eq. \eqref{fprime}. The above expression is not sign-definite but depends on the properties of the solution. For instance, for adequately large values of the throat radius, the above is clearly positive. We note therefore that any effective violation of the Null Energy Condition does not necessarily imply a negative $\rho_{\mbox{\tiny{eff}}}$. The violation may be caused merely by the negative radial pressure, which is the one that keeps the throat open -- indeed, from Eqs. \eqref{Gtt} and \eqref{Grr}, we readily obtain that at the throat the anisotropic pressure component is $$(p_r)_{\mbox{\tiny{eff}}}|_{r_0}=-\frac{1}{r_0^2} <0.$$  

%In  of the theory.} 

Let us now consider geodesic motion in the equatorial plane of the wormhole. Our analysis is not exhaustive -- we only seek to point out important  differences as compared to a standard black hole geometry. For a start, consider timelike geodesics of some test particle with rest mass $m$. We have that
\begin{equation}
\label{time}
    -h\dot{t}^2+\frac{W}{h}\dot{r}^2+r^2 \dot{\theta}^2+r^2 \sin{\theta}^2 \dot{\varphi}^2=-m^2
\end{equation}
where an overdot denotes the derivative with respect to proper time. We consider equatorial geodesics, $\theta=\pi/2$, and our Killing symmetries yield
\begin{equation}
 r^2 \dot{\varphi}=L,\qquad h  \dot{t}=E\,,
 \label{cons}
\end{equation}
where $L$ and $E$ are the conserved angular momentum and rest energy of the particle. The above three, first order, differential equations describe in full the equatorial timelike geodesics. Combining (\ref{time}-\ref{cons}) we obtain an effective Schrodinger like equation,
\begin{equation}
    (\sqrt{W} \dot{r})^2+\left(1+\frac{\tilde{L}^2}{r^2}\right)h=\tilde{E}^2\,,
    \label{schro}
\end{equation}
where we have divided by $m^2$ and thus the tilded  quantities are now given per unit rest mass (as is the curve parameter).
The first term of the LHS of (\ref{schro}) can be interpreted as the kinetic energy and the second term as the effective potential, $V_{eff}=\left(1+\frac{\tilde{L}^2}{r^2}\right)h$, while on the RHS we have the conserved overall energy $\tilde{E}^2$. This equation reduces to the usual timelike geodesic equation for a Schwarschild black hole if we set $W=1$ and $h=1-2M/r$, as discussed in standard GR textbooks. The difference here occurs essentially due to the wormhole function $W$ which changes the Schrodinger variable to $dR=\sqrt{W} dr$. As a result, the zeros of kinetic energy (and therefore the extrema of the potential) include now the throat location at $r=r_0$. In other words, the throat location is now a turning point or a point of equilibrium for geodesic motion parametrised by $\tilde{L}, \tilde{E}$, the initial data for the test particle. In fact for $r=r_0$ we have a circular orbit for the wormhole geometry for given $E_0$ and $L_0$,
\begin{equation}
    \tilde{E_0}^2=(1-\lambda)^2\left(1+\frac{\tilde{L_0}^2}{r_0^2}\right)
\end{equation}
while $$V_{eff,R}(r_0)=\sqrt{W^{-1}}(r_0) V_{eff,r}(r_0)=0.$$ The sign of the second derivative tells us if the trajectory is stable or not. This is a clear difference of the wormhole throat with respect to the absence of the black horizon surface \cite{Damour:2007ap} absorbing all in-falling matter. 
%The $\alpha$ goes to zero.

\begin{figure}[t] 
%\lbfig{Fig_phi} 
\begin{center}
\hspace{0.0cm} \hspace{-0.3cm}
\includegraphics[height=.22\textheight, angle =0]{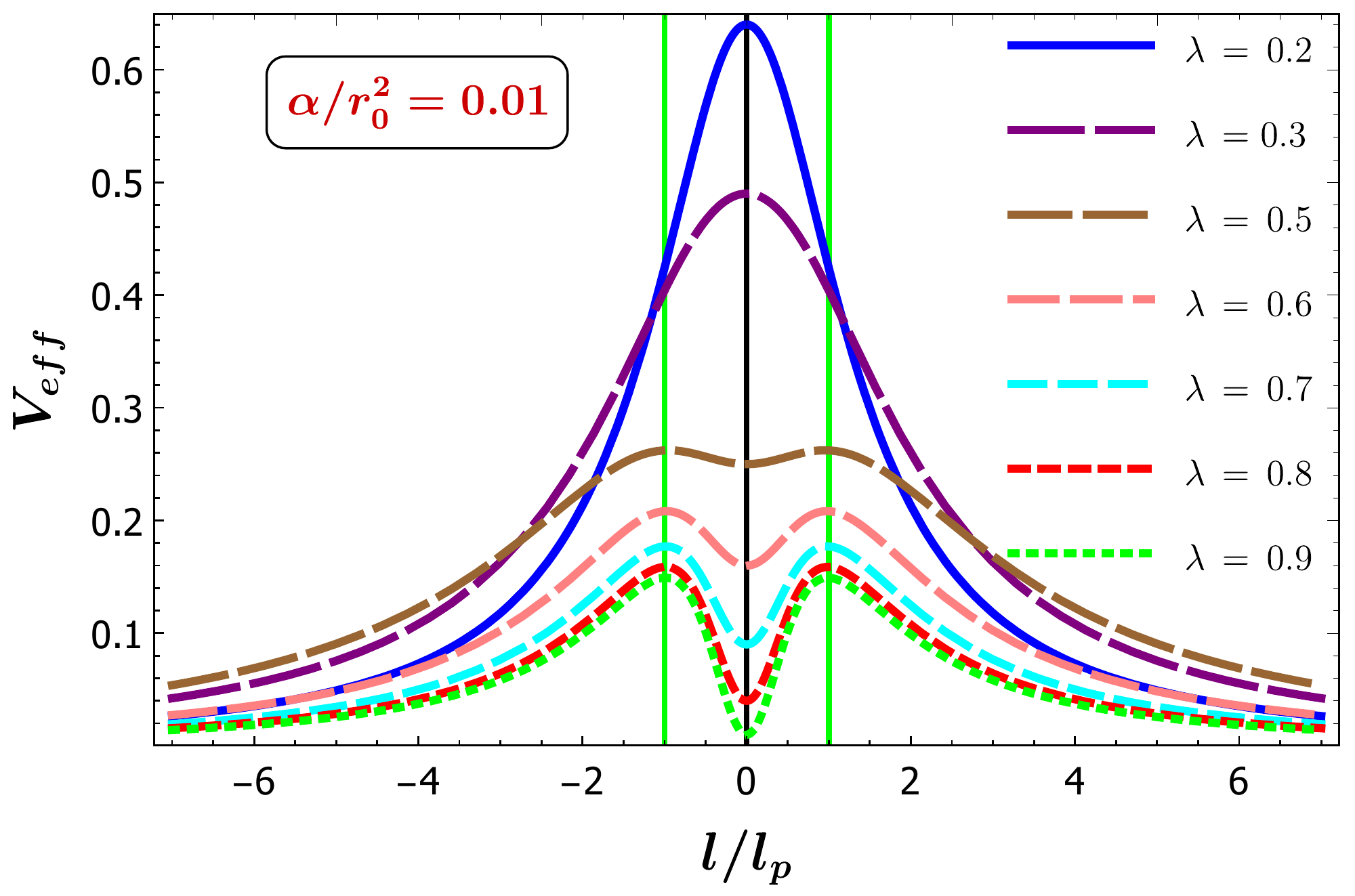}
\hspace{0.3cm} \hspace{-0.5cm}
\includegraphics[height=.22\textheight, angle =0]{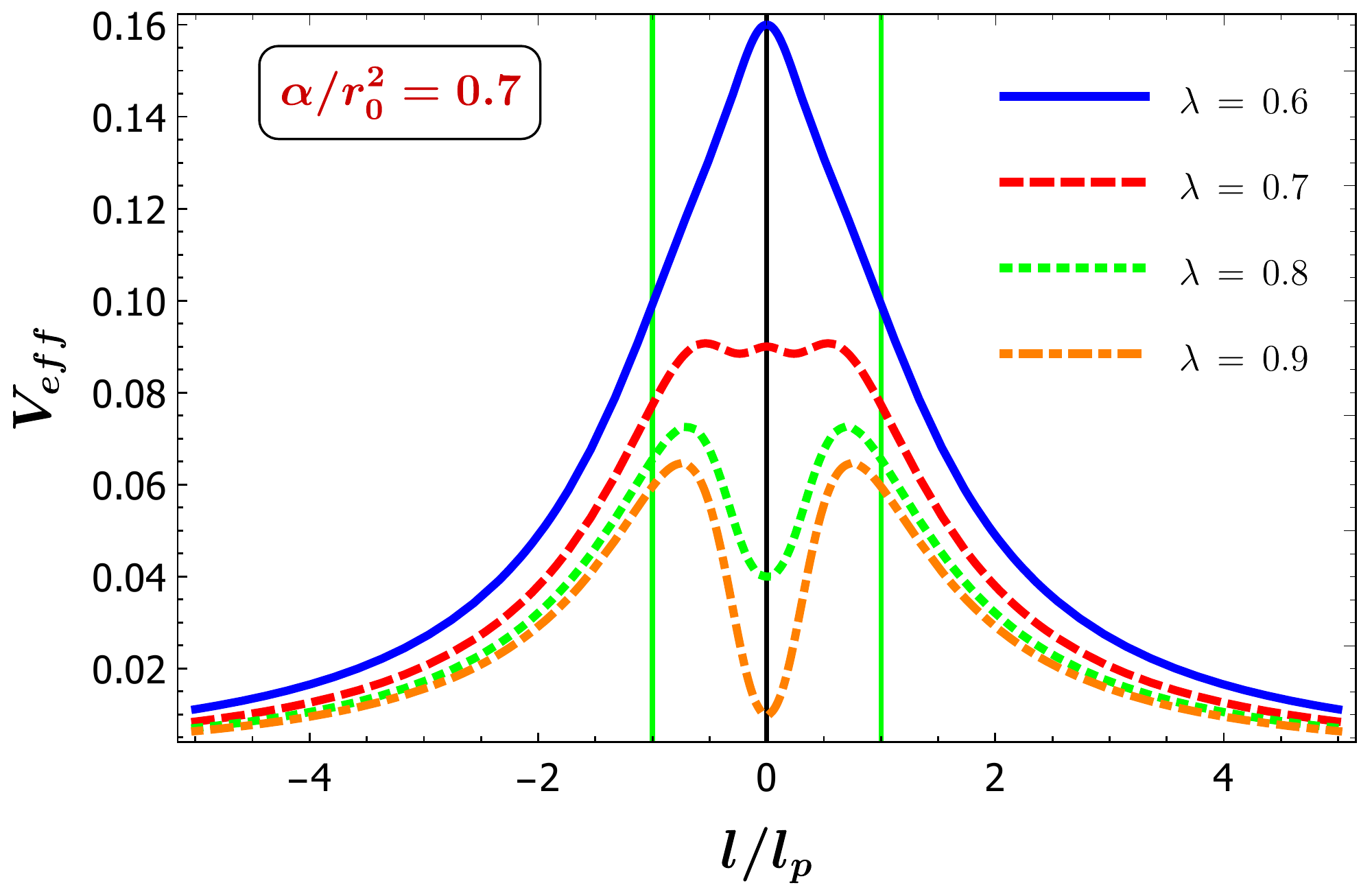}
\\
\hspace*{0.7cm} {(a)} \hspace*{7.5cm} {(b)}  \vspace*{-0.1cm}
\hspace{0.2cm} \hspace{-0.5cm}
\end{center}
\caption{The effective potential $V_{eff}$ for light geodesics as a function of the coordinate $l$ normalized at $l_p=\sqrt{|(3M)^2-r_0^2|}$.}
  \label{veff}
\end{figure}

We can also investigate briefly light geodesics by defining the impact parameter $b=L/E$. In a similar fashion we obtain,
\begin{equation}
    W \dot{r}^2+\frac{h}{r^2}=\tilde{b}^{-2}\,,
\end{equation}
where the overdot is now a derivative with respect to an affine parameter, and $$V_{eff}(R)=\frac{h(r(R))}{r(R))^2}.$$ The possible light rings (or circular light orbits) associated to this potential are again given by the extrema of $V_{eff}$, but crucially, as a function of $R$ (and not $r$ as for the black hole). Indeed we have that,
$$ V_{eff,R}=\sqrt{W^{-1}} \left( \frac{h(r)}{r^2}\right)',$$ therefore the throat is always an additional critical point. To get a qualitative idea of the possible critical points, we start with the case of $\alpha\rightarrow0$, whereupon $h$ is close  to Schwarzschild. We have therefore two possible extrema, one  at $r=3M$ and one at the throat, $r=r_0$. If $1>\lambda>1-\frac{1}{\sqrt{3}}$, we get $r_0<3M$, and the throat is a stable circular orbit while $r=3M$ is the unstable light ring [see, for example, the solutions with $\lambda\geq0.5$ in Fig. \ref{veff}(a)]. If $\lambda< 1-\frac{1}{\sqrt{3}}$, the throat is an unstable circular orbit at $r_0>3M$ replacing the (usual) light ring which is now excised from spacetime [see the solutions with $\lambda\leq0.3$ in Fig. \ref{veff}(a)]. 

For the general case with $\alpha\neq 0$, solving for $\frac{dV_{eff}}{dr}=0$ we get a third order polynomial,
\begin{equation}
    r^3-9M^2 r+ 8 \alpha M=0\,,
    \label{rings}
\end{equation}
which can have up to three roots, two at most of which turn out to be positive. Again starting with the familiar $\alpha \rightarrow 0$ case, we have that the extrema are at $r=0$ and $r=3M$. As we gradually increase $\alpha$, the two extrema $(r_{ext}^1, r_{ext}^2)$, initially at $(0,3M)$, start to approach each other satisfying $0<r_{ext}^1\leq r_{ext}^2<3M$, up to the point where they become one at $\alpha_{crit}=\frac{3\sqrt{3}M^2}{4}${\footnote{Note that the seed metric is already singular at such $\alpha$ as $\alpha_{crit}>M^2$.}}.
Summing it all up, we have up to three possible critical points depending on the throat position with respect to the two possible zeros of (\ref{rings}). This is the case if $\lambda$ is sufficiently close to unity and $\alpha$ is quite large so that $r_0<r_{ext}^1< r_{ext}^2$. In this case, the wormhole light structure (see also \cite{Cardoso:2016rao}) is quite intriguing with two unstable and one stable light ring in-between [see, for example, the curve for $\lambda=0.7$ in Fig. \ref{veff}(b)].

We can also evaluate the time it takes for a light ray to cross the wormhole throat (see for example \cite{Damour:2007ap}, \cite{MT}), as measured by an observer residing far away from the throat. Let us consider a point close to the throat, $l=0$, say at $l=l_0$. We have that
\begin{eqnarray}
    \Delta t&=&2\int_0^{l_0} \frac{d \xi}{\sqrt{h(\xi)}}=2\int_0^{l_0} \frac{dl}{\sqrt{H(l)F(l)}}=\nonumber\\
    &=&\frac{1}{\sqrt{h_0 f_0}}\left[2 l_0-\frac{l_0^3}{3}\left(\frac{h_2}{h_0}+\frac{f_2}{f_0} \right)\right].
\end{eqnarray}
Here we are making use of the regular $l$ coordinates as well as the expansions in Appendix B{\footnote{Proper distance coordinate, $\xi$, agrees with $l$ coordinates for small enough $l$ as, $\xi=l+0(l^3)$}}. Thus, the time interval $\Delta t$ for such a crossing depends once again on the particular values of the parameters of the theory.

%Note for $r_h<r_0<r_p$.

%%%%%%%%%%%%%%%%%%%%%%%%%%%%%%%%%%%%%%%%%%%%%%%%%%%%%%%%%%%%%%%%%%%%%%
Let us now turn to the acceleration and tidal forces a potential observer would feel while traversing the wormhole. Following the analysis of \cite{MT, Kanti:2011yv} and performing appropriate coordinate transformations, we may change our reference frame first to that of a static observer at an arbitrary radial distance $\ell$ and then to the frame of the moving traveler with basis vectors $(\hat e_{\tilde t}, \hat e_{\tilde \ell}, \hat e_{\tilde \theta}, \hat e_{\tilde \varphi})$. The acceleration felt by the traveler satisfies the relation $a^{\tilde\mu} =u^{\tilde \nu}\,\nabla_{\tilde \nu} u^{\tilde \mu}$, where $u^{\tilde \mu}$ is the normalised four-vector velocity, i.e. $u^{\tilde \mu} u_{\tilde \mu} = -1$. If we assume that the traveler moves radially, we have $a^{\tilde \mu}=a \hat e_{\tilde \ell}$. Working backwards, we may express the magnitude of the acceleration $a$ in terms of quantities employed by the asymptotic observer, and thus write
\begin{equation}
    |a|=\sqrt{\frac{F}{H}}\,\partial_\ell \Bigl( \gamma \sqrt{H}\Bigr)\,,
\end{equation}
where $\gamma=1/\sqrt{1-u^2/c^2}$. If the traveler moves with non-relativistic velocity, i.e. $u \ll c$, we may set $\gamma \simeq 1$, in the above expression. By changing back to the radial coordinate $r$, the above is written as
\begin{equation}
    |a| = W^{-1/2} \,\partial_r \sqrt{h}= \sqrt{1-\frac{r_0}{\lambda r}\,(1-\sqrt{h})}\,\,\partial_r \sqrt{h}\,.
\end{equation}
Far away from the throat, both $W^{-1}$ and $h$ reduce to unity and the acceleration vanishes as expected. But also at the location of the throat, where $W^{-1}$ again vanishes, the acceleration measured is zero. In the intermediate regime, the magnitude $|a|$ depends strongly on the parameter $\lambda$. As $\lambda$ decreases, and the radius of the throat $r_0$ increases, the acceleration is suppressed. 

We may also calculate the tidal force felt by the observer between two parts of their body. In the traveler's reference frame, their separation vector is $k^{\tilde \mu}=(0, k^{\tilde i})$. Following again the analysis in 
\cite{MT, Kanti:2011yv}, we find that the radial and transverse components of the tidal force are
\begin{eqnarray}
\Delta a^{\tilde \ell}&=& c^2 k^{\tilde \ell} F
\left( -\frac{F' H'}{4F H} + \frac{H'^2}{4 H^2} - \frac{H''}{2H}\right)\,, \\[1mm]
\Delta a^{\tilde \theta}&=& \frac{c^2 \gamma^2 k^{\tilde \theta}}{(\ell^2+r_0^2)} \left[-\frac{\ell F H'}{2H} + \frac{u^2 F}{c^2}\left(\frac{r_0^2}{\ell^2+r_0^2}  + \frac{\ell F'}{2F}\right)\right]\,.
\end{eqnarray}
As was the case with the acceleration, both components of the tidal force vanish in the asymptotic limit as $H$ and $F$ reduce to unity and $\ell \rightarrow \infty$. Near the throat, they both acquire constant values given by the expressions
\begin{equation}
 \Delta a^{\tilde \ell} \simeq - c^2 k^{\tilde \ell}\, \frac{f_0 h_2}{h_0}\,, \qquad \Delta a^{\tilde \theta} \simeq \frac{\gamma^2 u^2 k^{\tilde \theta} f_0}{2 r_0^2}\,,
\end{equation}
where we have used the expansions of the metric functions near the throat given in Appendix B. It is straightforward to see that, as we decrease $\lambda$, and thus increase $r_0$, both components of the tidal force are suppressed since $f_0 \rightarrow 1$, $h_0 \rightarrow 1$ and $h_2 \rightarrow M/r_0^3$.

We have shown in the previous section that the wormhole solutions are everywhere regular. This signifies that curvature scalars are always finite. They are however indicative for the traversability of the wormhole. 
Let us consider the Ricci scalar, 
%%%%
\begin{equation}
   R=\frac{W' h'}{2 W^2}+\frac{2 h W'}{W^2 r}-\frac{h''}{W}-\frac{4 h'}{W r}-\frac{2
   h}{W r^2}+\frac{2}{r^2}\,.
\end{equation}
%%%%%%
Focusing in particular on its behaviour at the throat, we obtain
%%%%%%%%
\begin{equation}
    R(r_0)=\frac{2}{r_0^2} -\frac{2 f'(r_0)}{r_0} +\frac{W' h'}{2 W^2}\bigg|_{r_0}\,.
    \label{R_throat}
\end{equation}
%The  the wormhole.

%%%%%%%%%%%%%%%%%%%%%%%%%
%
\begin{figure}[t] 
%\lbfig{Fig_phi} 
\begin{center}
\hspace{0.0cm} \hspace{-0.3cm}
\includegraphics[height=.22\textheight, angle =0]{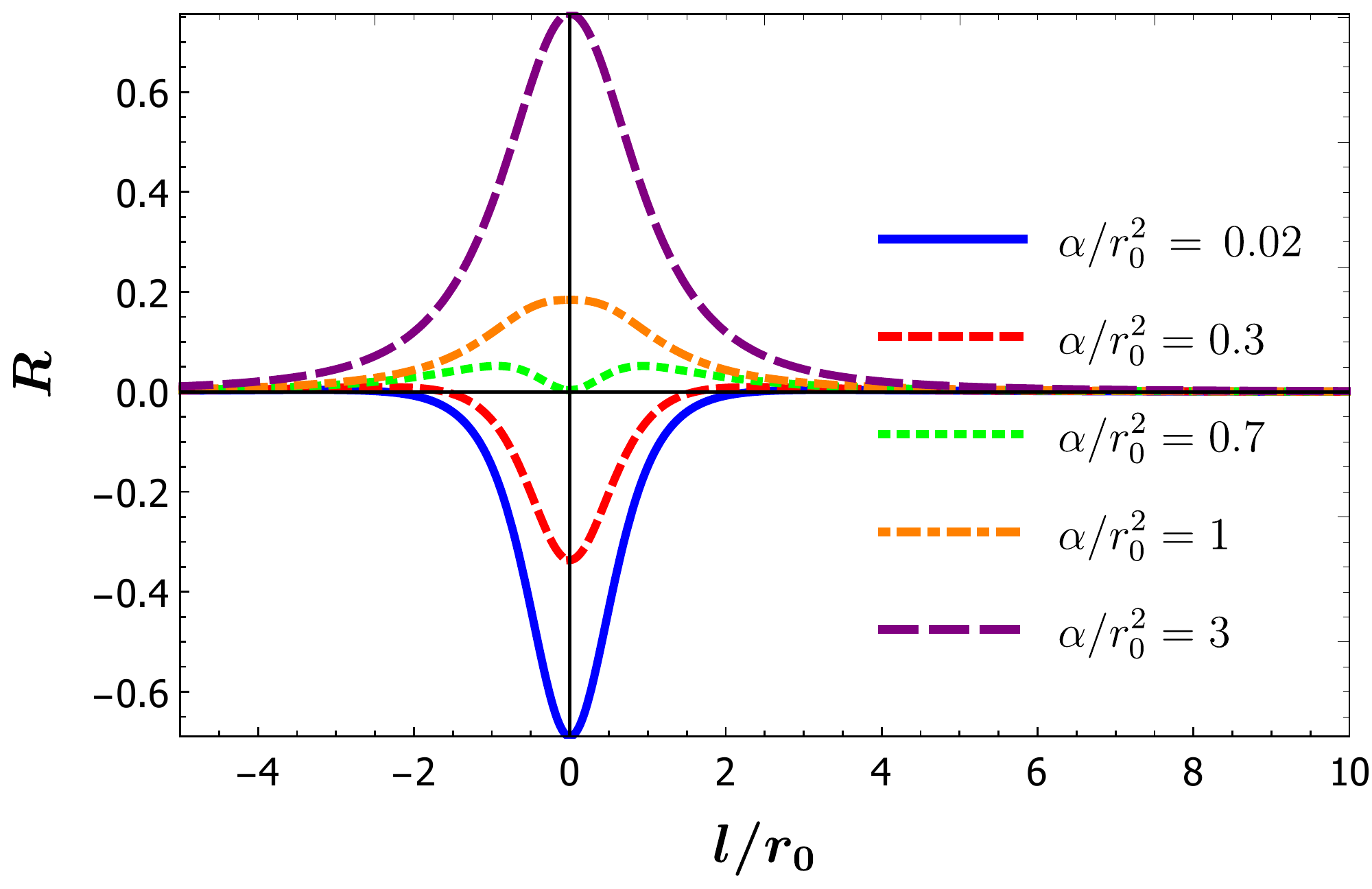}
\hspace{0.3cm} \hspace{-0.5cm}
\includegraphics[height=.22\textheight, angle =0]{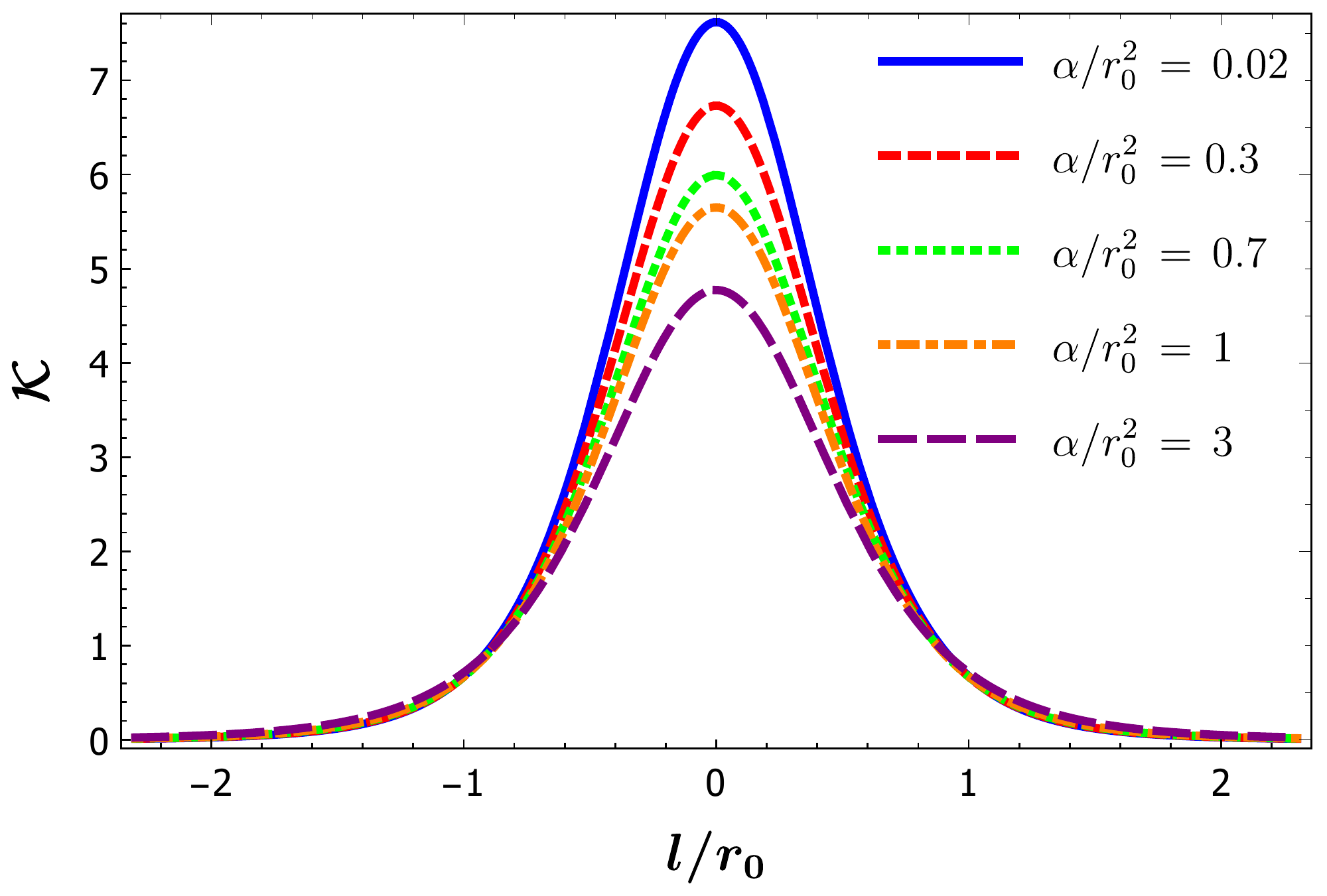}
\\
\hspace*{0.7cm} {(a)} \hspace*{7.5cm} {(b)}  \vspace*{-0.1cm}
\hspace{0.2cm} \hspace{-0.5cm}
\end{center}
\caption{The  Ricci scalar $R$ and Kretchmann scalar ${\cal K}=R_{\mu\nu\rho\sigma} R^{\mu\nu\rho\sigma}$ for a number of wormhole solutions with $\lambda=0.2$.}
  \label{Ricci}
\end{figure} 
%
%%%%%%%%%%%%%%%%%%%%%%%%%%

%The regularity. 
In Fig. \ref{Ricci}, we depict the Ricci scalar $R$ and the Kretchmann scalar ${\cal K}=R_{\mu\nu\rho\sigma} R^{\mu\nu\rho\sigma}$ for a number of wormhole solutions with fixed $\lambda=0.2$. Both quantities,
%-- as well  not shown here -- 
being regular over the entire spacetime, render the wormhole {\it traversable} by particles. We also observe that the Ricci scalar is not sign-definite at the throat but depends on the particular solution.  As Fig. \ref{Ricci}(a) reveals, wormhole solutions with small values of the parameter $\alpha/r_0^2$ have a negative Ricci scalar at the throat whereas, as $\alpha/r_0^2$ increases, $R$ may adopt either a vanishing or a positive value at this point. 

%The $R$ at the throat.  

%%%%%%%%%%%%%%%%%%%%%%%%%%
\begin{figure}[b!] 
%\lbfig{Fig_phi} 
\begin{center}
\hspace{0.0cm} \hspace{-0.5cm}
\includegraphics[height=.215\textheight, angle =0]{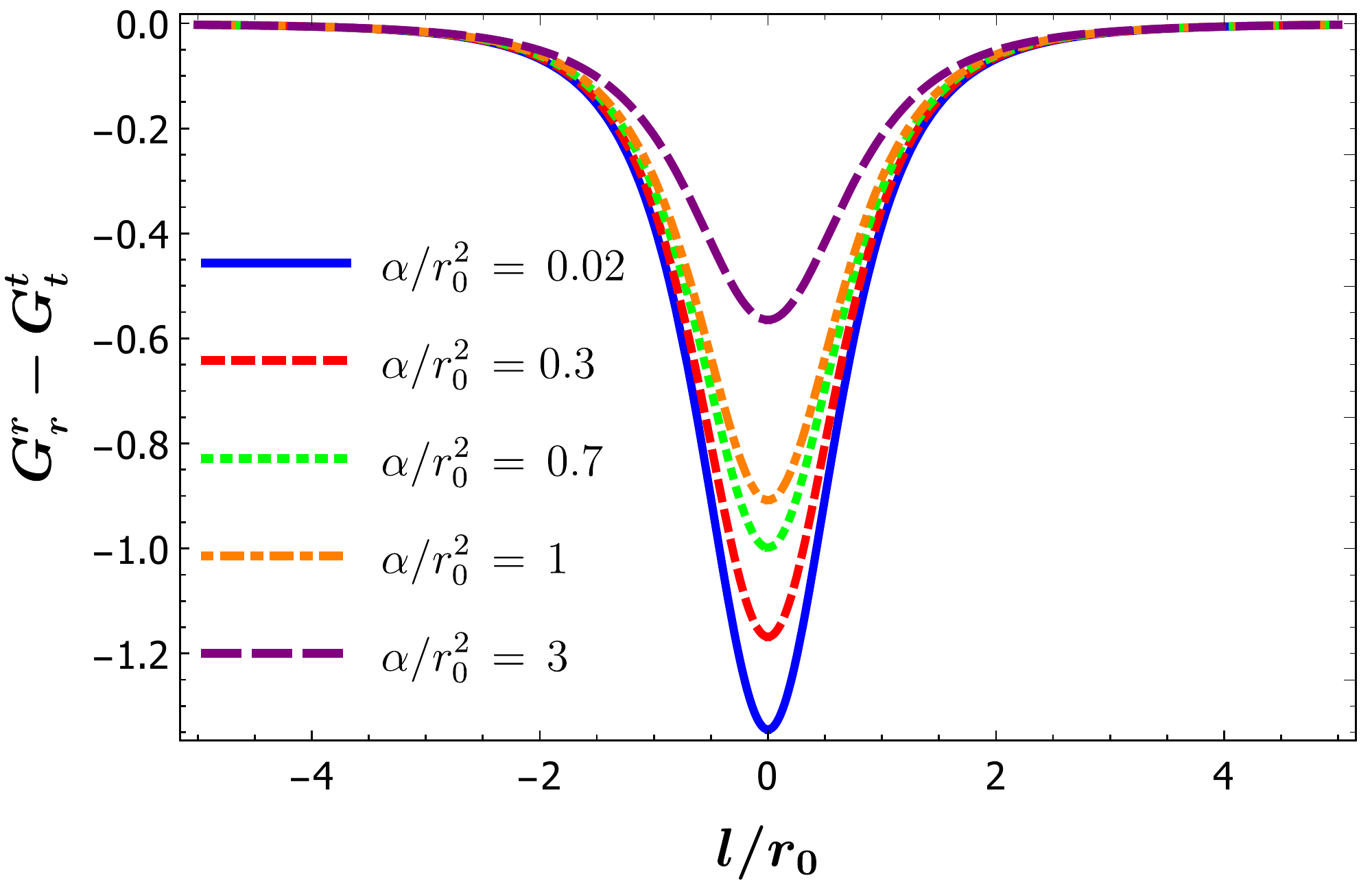}
\hspace{0.22cm} \hspace{-0.3cm}
\includegraphics[height=.21\textheight, angle =0]{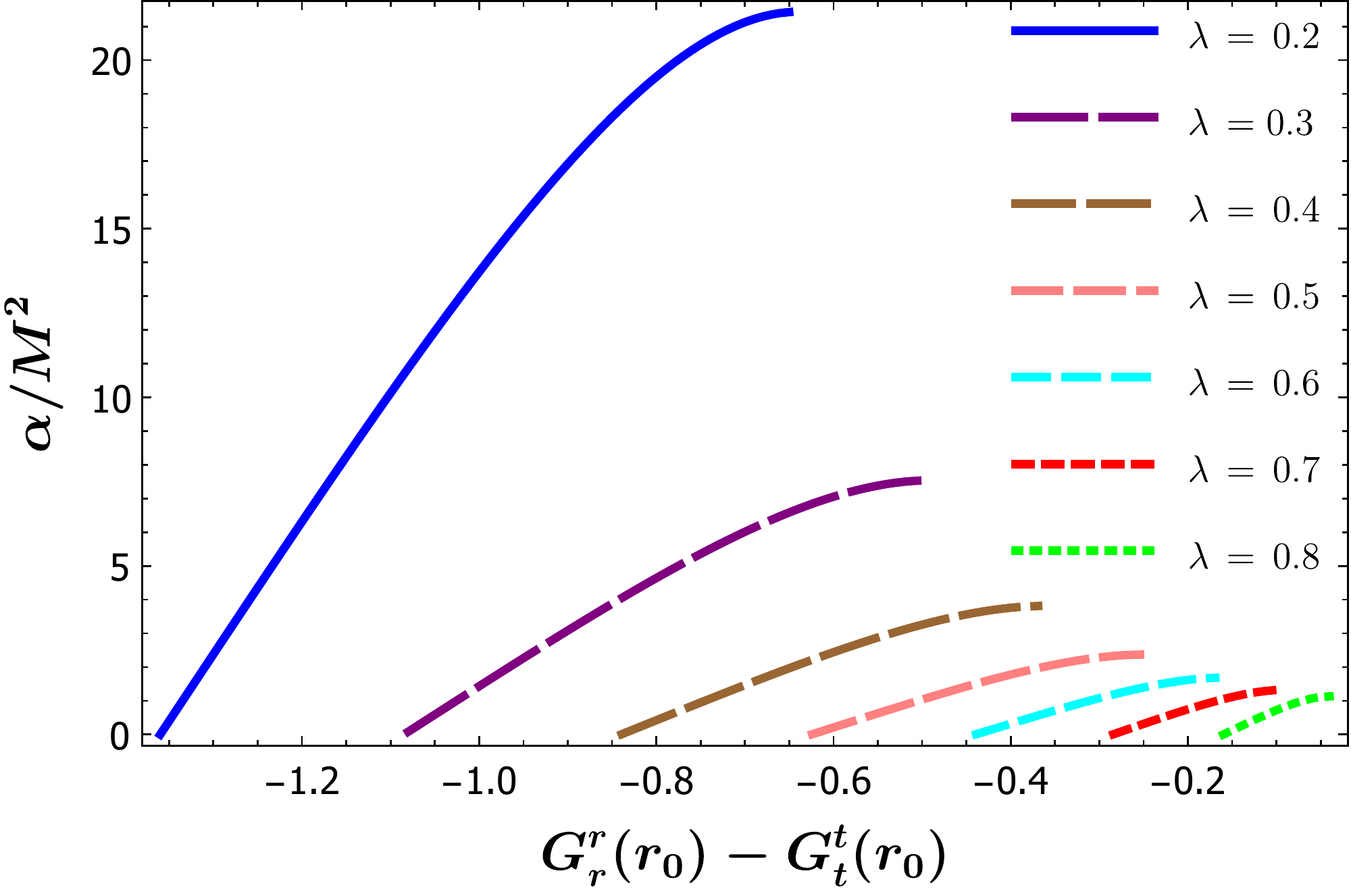}
\\
\hspace*{0.7cm} {(a)} \hspace*{7.2cm} {(b)}  \vspace*{-0.5cm}
\end{center}
\caption{(a) The first NEC energy condition $G_r^r-G_t^t$ in terms of the coordinate $l$, and (b) its value at the throat in terms of the parameter $\alpha/M^2$ and for different values of $\lambda$.}
  \label{neq1}
\end{figure} 
%%%%%%%%%%%%%%%%%%%%%%%%%%%%%%%%

In Fig. \ref{neq1}(a), we display the first NEC energy condition $T_r^r-T_t^t$ in terms of the radial coordinate $l$, and for a number of our wormhole solutions parametrised again by the value of $\alpha/r_0^2$. We observe that the first NEC energy condition is always violated at the throat, and this violation persists throughout the spacetime. As we discussed at the beginning of the section,  the violation of the first NEC at the location of the throat is a prerequisite for the existence of the throat itself. The same holds for all of the wormhole solutions we have found in our analysis. In Fig. \ref{neq1}(b), we present the domain of existence of our solutions, i.e. the first NEC condition at the throat for the complete parameter range of $\alpha/M^2$ and for several different values of the disformal parameter $\lambda$. We observe that, as the value of $\lambda$ increases, the allowed range of values of $\alpha/M^2$ decreases - this readily follows from the bound of Eq. (\ref{alpha_bound}). As is expected, the violation of the first NEC is always realised, for the wormhole geometry to emerge, however, this violation gets milder as $\lambda$ increases.

%%%%%%%%%%%%%%%%%%%%%%%
%
\begin{figure}[t!] 
%\lbfig{Fig_phi} 
\begin{center}
\hspace{0.2cm} \hspace{-0.5cm}
\includegraphics[height=.21\textheight, angle =0]{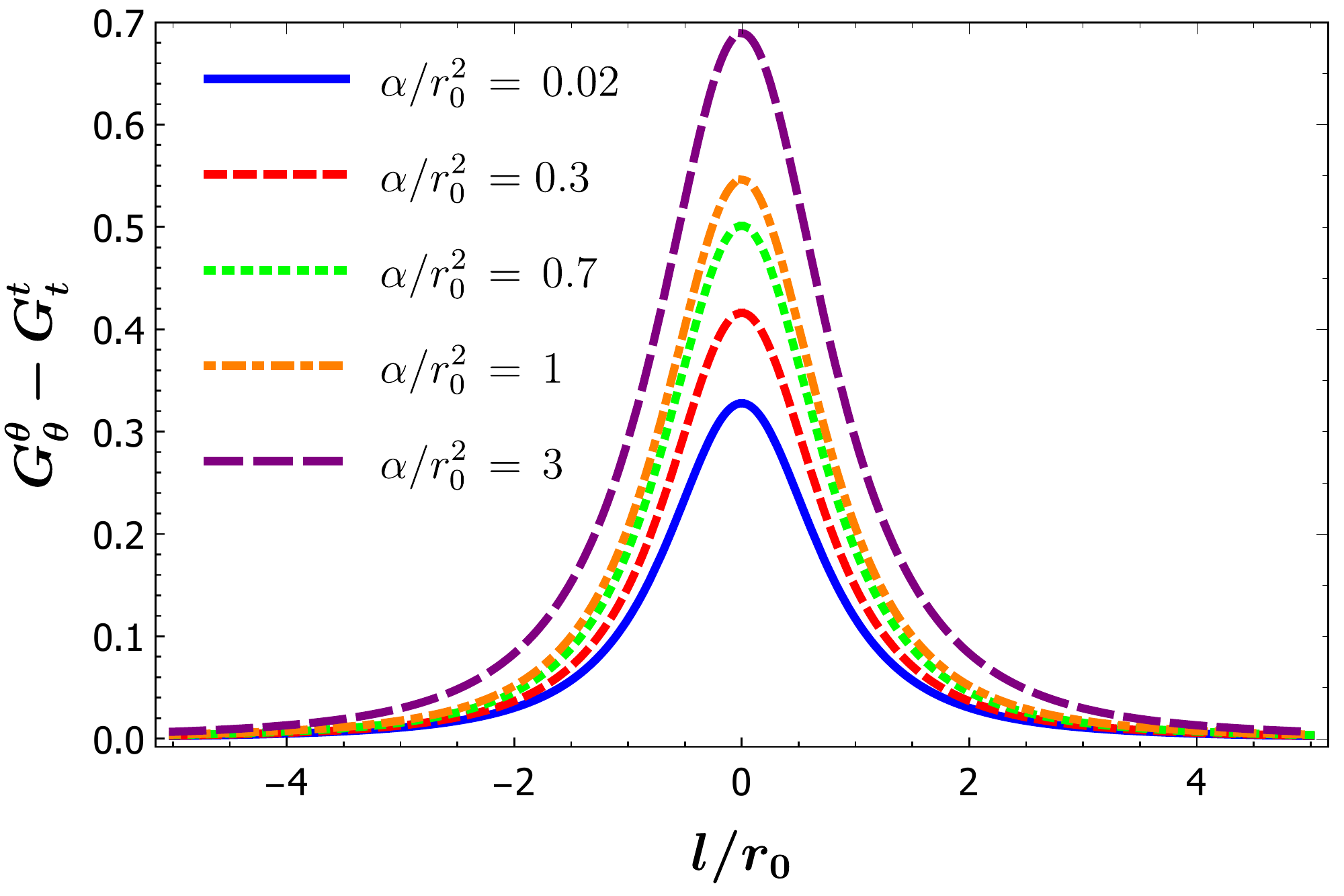}
\hspace{0.22cm} \hspace{-0.3cm}
\includegraphics[height=.21\textheight, angle =0]{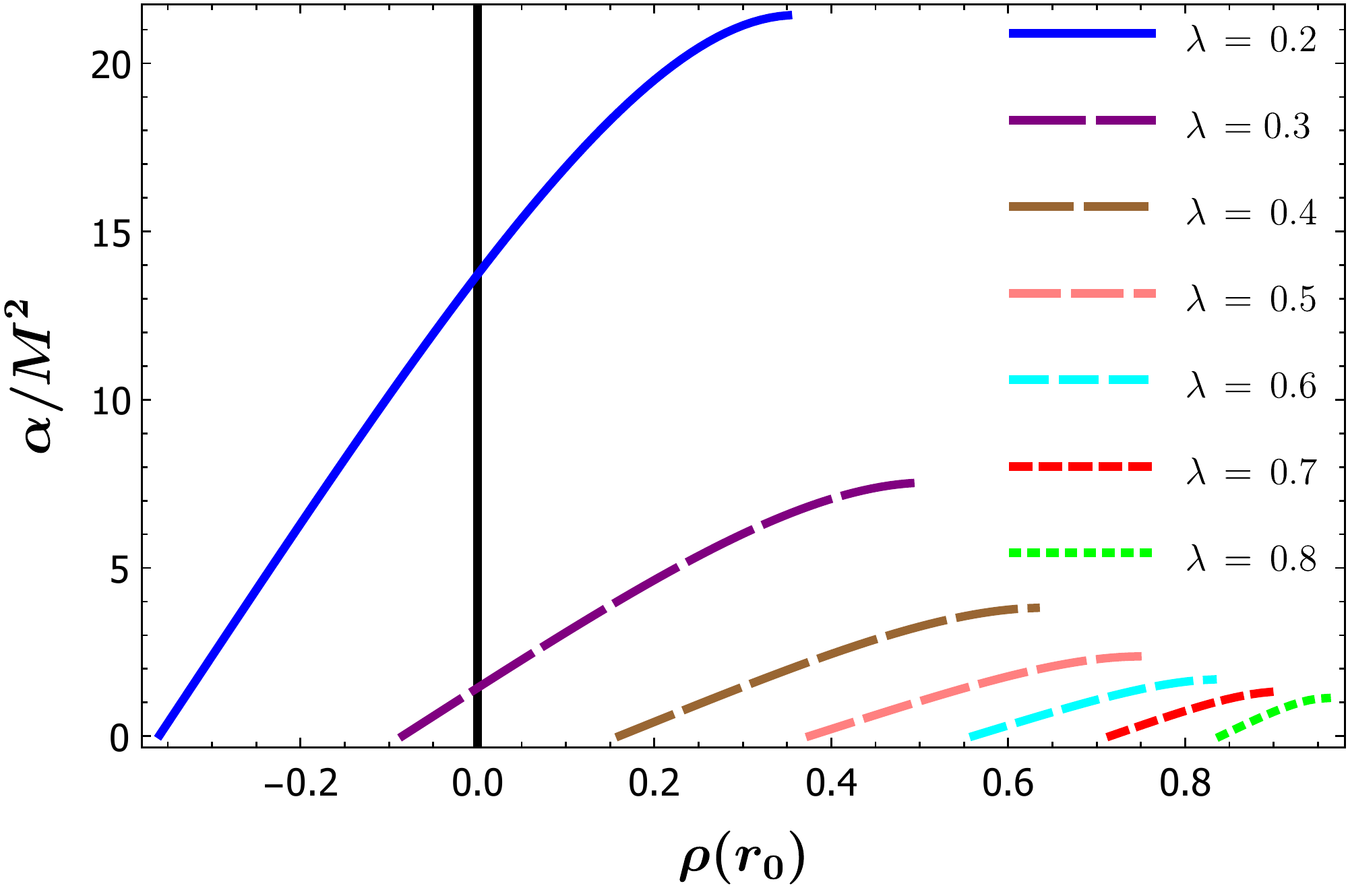}
\\
\hspace*{0.7cm} {(a)} \hspace*{7.2cm} {(b)}  \vspace*{-0.5cm}
\end{center}
\caption{(a) The second NEC energy condition $G_\theta^\theta-G_t^t$ in terms of the coordinate $l$, and (b) its value at the throat in terms of the parameter $\alpha/M^2$ and for different values of $\lambda$.}
  \label{sec}
\end{figure} 
%%%%%%%%%%%%%%%%%%%%%%

The second NEC energy condition $T_\theta^\theta-T_t^t$ is displayed in Fig. \ref{sec}(a), and is clearly obeyed over the entire spacetime since it remains always positive. As was discussed earlier, this condition is not necessarily violated for a wormhole throat to emerge. Since it involves the effective energy density $\rho$ and transverse pressure component $p_\theta$, its non-violation allows for the possibility that solutions with $\rho>0$, which respect the WEC, may be found. In Fig. \ref{sec}(b), we thus present the domain of existence of our solutions where now the horizontal axis displays the energy density at the location of the throat $\rho(r_0)$. The depicted behaviour resembles the one for the Ricci scalar $R$ discussed above. For small values of $\alpha/M^2$ and small $\lambda$, only solutions with $\rho(r_0)<0$, which violate the WEC, are found. However, as either of these two parameters increases, we obtain wormhole solutions with {\it exclusively positive} $\rho(r_0)$, which therefore respect the WEC. We remind the reader that the solutions with $\rho>0$ constructed in \cite{MT} had the distribution of matter extending over the entire spacetime. Our subclass of wormholes with $\rho>0$ are in contrast {\it localised}: as Fig. \ref{Ricci} reveals, the curvature of the spacetime around all of our solutions reduces to very small values at distances $|l| \simeq 2 r_0$ and literally vanishes at $|l| \simeq 4 r_0$. Therefore, our solutions comprise a family of regular, traversable, localised wormholes respecting the WEC.

%%%%%%%%%%%%%%%%%%%%%%%%%%%%%%%%%%%%%%%%%%%%%%%%%%%%%%%%%%%%%%%%%%%
%
%
%%%%%%%%%%%%%%%%%%%%%%%%%%%%%%%%%%%%%%%%%%%%%%%%%%%%%%%%%%%%%%%%%%%

\section{Discussion and Conclusions}
\setcounter{equation}{0}

Horndeski theory \cite{Horndeski:1974wa}, the most general scalar-tensor theory of gravity leading to field equations with only up to second-order derivatives of the scalar field and metric tensor, has led to a number of black-hole solutions in the literature (see for example \cite{Lehebel:2018zga}, \cite{Anson:2021msx}). Here, we have turned our attention to the construction of wormholes, and formulated a method to derive such solutions by employing a disformal transformation of the metric tensor: starting from a well-known black-hole solution with a scalar field, namely the Lu-Pang solution \cite{Lu:2020iav}, and choosing appropriately the disformal function, we have transformed the background metric to a new form which is characterised by the presence of a throat and not an event horizon (see also the very recent works \cite{Faraoni:2021gdl} and \cite{Chatzifotis:2021hpg}). The new theory at hand, after the disformal transformation, is in beyond Horndeski theory. An important point concerning our seed solution is that the kinetic scalar term $X$ is not constant, otherwise the disformed solution is anew a black hole with rescaled parameters (see for example \cite{Babichev:2017lmw}). 

The throat corresponds to a root of the shape function of the metric, i.e. of the $g^{rr}$ component, and by construction is located at a radial distance $r_0$ larger than the seed horizon radius $r_h$ of the original Lu-Pang black hole. We note that the presence of the horizon for the seed metric is actually not necessary; in this case the throat must be situated to the right of the local minimum of the seed metric component (rather than the horizon radius). The $g_{tt}$ component is strictly  negative throughout the wormhole spacetime, playing the role of the red-shift function of the wormhole geometry. The smaller the $g_{tt}$ component is, at the throat position, the closer the wormhole simulates the geometry of a black hole. 
%We scalar field.   

By changing to a new coordinate system, we excise the irrelevant part of the seed geometry and obtain a completely regular chart of coordinates. We have thus constructed the complete regular wormhole geometry comprising two asymptotically-flat regions and a regular throat located at the origin of the new spatial coordinate. Both metric functions remain everywhere regular (of at least $C^2$ regularity), and exhibit a smooth transition from the positive to the negative radial coordinate, a result which renders unnecessary the presence of any additional matter distribution at the wormhole throat. The curvature invariants confirm that the spacetime is singularity-free, and our wormhole solutions are therefore traversable by particles. They also reveal that certain wormholes can be classified as localised solutions since the curvature of spacetime virtually vanishes at distances only a few times the wormhole throat. In our parametrization of the solutions these cases occur whenever the $\lambda\in(0,1)$ parameter is close to unity. This case is close to the geometry and properties described in the paper by Damour and Solodukhin \cite{Damour:2007ap}. In fact, when $\lambda=1$ the throat becomes an event horizon and the throat is a one-way membrane. This of course as long as the Horndeski or Lovelock parameter $\alpha$, present in the seed metric, is such that $\alpha<M^2$, guarantying the presence of an event horizon for the seed metric \cite{Lu:2020iav}. When, on the other hand, $\lambda$ is closer to zero, the wormhole approaches the zero tidal force type wormhole described in the classic Morris-Thorne paper \cite{MT}.

The disformal transformation leaves in principle unchanged the form of the scalar field, however, its actual configuration changes implicitly as it depends on the gravitational background. In the absence of a horizon or a singularity, the scalar field is also everywhere regular: it takes its lowest value at the wormhole throat and approaches a constant value at the two asymptotic infinities -- the latter conforms with the localisation of all energy-momentum components near the wormhole throat and the emergence of the two asymptotically-flat regimes. 

%The  energy-density. 

The construction of the aforementioned wormhole solutions was realised by choosing a particular disformal function given in Eq. (\ref{disformal}). By applying this to the Lu-Pang metric tensor, we have changed not only the background metric but also the theory itself. The disformal transformation takes a Horndeski solution to a beyond Horndeski one. The form of the coupling functions of the new theory are given in Appendix A.  As already stated, this is not the only choice of disformal function one could make. For example, one could consider the general form
\begin{equation}
    W^{-1}=1+\frac{1}{\lambda}\sum_{i=1}^n c_i\left( r_0 \sqrt{-2\bar X}  \right)^i\,,
\end{equation}
where $c_i$ and $\lambda$ are dimensionless parameters. 
Our choice (\ref{disformal}) constitutes a particular case of the above general expresion for $n=1$ and $c_1=1$.  We have confirmed that wormhole solutions with the same set of attractive characteristics emerge also for a number of particular forms of the above disformal function. For example, for $n=2$, $c_1=-2$, and $c_2=1$, we obtain
%%%%%
\begin{equation}
W(\bar X(r_0))^{-1}=1+\frac{h(r_0)-1}{\lambda}\,.
\end{equation}
%%%%
Demanding that the above vanishes at the location of the throat leads to the constraint $h(r_0)=1-\lambda$. Using again the original Lu-Pang metric function $h(r)$, we find that in this case the throat is located at
%%%%%%
\begin{equation}
     r_0=\frac{M\pm \sqrt{M^2- \alpha \lambda^3}}{\lambda}\,,
\end{equation}
which is actually a simpler expression compared to the one of Eq. (\ref{roots}). One could study these in greater detail to see if there are any other additional effects from those described here. 
%We refrainady presented.

The construction method for wormhole solutions proposed in this work can be applied to any black-hole solution arising in the framework of Horndeski theory as long as they are not of the $X$-constant type{\footnote{Indeed if $X$ is constant, then the disformed metric, in the case of spherical symmetry at least, is the same solution with rescaled parameters \cite{Domenech:2019syf}, \cite{BenAchour:2020wiw}. This can have interesting consequences, e.g. for dark energy solutions \cite{Babichev:2017lmw}, but does not change the nature of the spacetime metric.}}. The resulting wormholes  are by construction solutions in beyond Horndeski theory. In fact, the method may be applied to other solutions of Horndeski theory, and not only on black holes, since the part of spacetime which contains the event horizon (or a naked singularity) is discarded and therefore irrelevant. It would be interesting to study the fate of black holes in beyond Horndeski theories where wormholes are constructed. Indeed, note that once the wormhole is constructed the seed black hole ceases to exist in the resulting theory and one would have to search for an alternative solution. Furthermore, the most promising filter to keep or discard solutions would seem to be the stability requirements (see \cite{Mironov:2018uou}, \cite{Volkova:2019kyd}) which are very strict in the case of Horndeski theory \cite{Evseev:2017jek}, \cite{Evseev:2018fma}. The construction of possible explicit rotating wormhole solutions (see for example \cite{Teo:1998dp}) would also be a direction worthy to follow; however, this seems at the moment quite a difficult task as the sole explicit rotating black holes are of the $X$-constant type. 
Last but not least, it would also be interesting to study in greater detail the geodesic motion of particles and light as well as the resulting shadows and astrophysical effects of the explicit solutions described here. 
Such a study could shed additional light in the differences (and similarities) regarding the observable signatures between black hole and wormhole geometries, and provide specific predictions which could serve as test points of modified theories of gravity. 

%We work. 

\section*{Acknowledgements}
We are very happy to thank Marco Crisostomi, Antoine Leh\'ebel, Karim Noui and Theodoros Nakas for encouraging and useful discussions. We last but not least thank Nicolas Lecoeur for his insightful remarks regarding the parametrisation of the wormhole solution and the corresponding theory.
A.B. and C.C. happily acknowledge networking support by the GWverse COST Action CA16104, “Black holes, gravitational waves and fundamental physics.” C.C. in particular thanks the Department of Physics in the University of Ioannina for hospitality during the course of this work.

%\newpage
\appendix
\renewcommand{\theequation}{\thesection.\arabic{equation}}
\addcontentsline{toc}{section}{APPENDIX\label{app}}

\section*{APPENDIX}

\section{The beyond Horndeski theory}
\setcounter{equation}{0}

We use $W^{-1}(\bar{X})=1-b_1\sqrt{-2\bar{X}}$ with $b_1=r_0/\lambda$, therefore the disformal transformation is
%%%
\begin{equation}
    D(\bar{X})=-\frac{b_1}{2b_1 \bar{X}+\sqrt{-2 \bar{X}}}.
\end{equation}
%%%%
For simplicity we use $\bar{X}=-y^2$. By substituting to Eqs. (\ref{g4d}-\ref{f5d}) and (\ref{g2d}-\ref{g3d}) we find
%%%

\begin{align}
 G_2&= 8\alpha y^4 (1-\sqrt{2}b_1 y)^{1/2}, \quad \quad  G_{3X}= 
 \frac{16 \alpha}{(3\sqrt{2}b_1 y-2)(1-\sqrt{2}b_1 y)^{1/2}},
 \\[3mm]
 G_4&= (1-4 \alpha  y^2) (1-\sqrt{2}b_1 y)^{1/2}, \quad\quad G_{5X}= -\frac{8 \alpha}{y^2 (3\sqrt{2}b_1 y-2)(1-\sqrt{2}b_1 y)^{1/2}}, 
   \\[3mm]
   F_4&= \frac{b_1 \left(1-2\sqrt{2}b_1 y\right) \left(4 \alpha  y^2+1\right)}{2^{3/2} y^3 \left(3 \sqrt{2} b_1 y-2\right) \left(1-\sqrt{2}b_1 y\right)^{5/2}},
   \\[3mm]
   F_5&=-\frac{\sqrt{2} \alpha b_1 \left(1-2\sqrt{2}b_1 y\right) }{3 y^5 \left(3 \sqrt{2} b_1 y-2\right) \left(1-\sqrt{2}b_1 y\right)^{7/2}} .
\end{align}
%%%%%
Finally from Eq. (\ref{fxd}) we find
\begin{equation}
    X=y^2(-1+\sqrt{2}b_1 y).
\end{equation}

\section{Near throat expansions}
\setcounter{equation}{0}

Near the throat of the wormhole (i.e. as $l\rightarrow 0$), the metric functions become
%%%%%%
\be
H(l) = h_0 + h_2\,l^2 + \mathcal{O}(l^4) \,, \qquad F(l) = f_0 + f_2\,l^2 + \mathcal{O}(l^4) \,,
\ee
%%%%%%
where
%%%%%%%
\begin{align}
\label{values1}
    h_0=&(1-\lambda)^2
    %h_0=&
    %\frac{2 \alpha -\sqrt{8 \alpha  M r_0+r_0^4}+r_0^2}{2 \alpha }
    ,\\[2mm]
    h_2=&\frac{1}{2\alpha}\,\left(1-\frac{2 \alpha  M+r_0^3}{r_0^{3/2} \sqrt{8 \alpha  M+r_0^3}}\right),\\[2mm]
    f_0=&\frac{(1-\lambda) \left[\alpha  \lambda  \left(3 \lambda ^2-8
   \lambda +4\right)+(4-3 \lambda ) r_0^2\right]}{4 \left(r_0^2-2
   \alpha  (\lambda -2) \lambda \right)},\\[2mm]
   f_2=&\frac{\lambda r_0 Q_1}{32 (\lambda -1) \left(r_0^3-2 \alpha  (\lambda -2) \lambda 
   r_0\right){}^3},\label{values4}
   \end{align}
with 
\begin{align}
   Q_1=&2 \alpha ^3 (\lambda -2)^3 \lambda ^2 (\lambda  (\lambda  (27 \lambda
   -76)+58)-12)\nonumber\\[2mm]
   -&3 \alpha ^2 (\lambda -2)^2 \lambda  (\lambda  (\lambda  (27 \lambda
   -92)+100)-32) r_0^2\nonumber\\[2mm]
   +&6 \alpha  (\lambda -2) (\lambda -1) (3 (\lambda -3) \lambda +10)
   r_0^4+(\lambda  (9 \lambda -32)+26) r_0^6\nonumber.
\end{align}
We note that the requirement $h_2>0$ results in the condition $r_0 > (\alpha M)^{1/3}=r_0^{min}$ derived in section 3 by following an alternative approach.

\bibliographystyle{utphys}
\bibliography{Bibliography}

\end{document}